\documentclass[useAMS,usenatbib,fleqn]{mn2e}

\usepackage{graphicx}
\usepackage{amssymb}
\usepackage{amsmath}
\usepackage{color}
\usepackage{ulem}

\newcommand{\aj}{AJ}
\newcommand{\apj}{ApJ}

\newcommand{\mnras}{MNRAS}
\newcommand{\nat}{Nature}
\newcommand{\newa}{New Astronomy}

\newcommand{\pasj}{PASJ}

\title[Binary black holes originating from globular
  clusters]{Dynamical evolution of stellar-mass black holes in dense
  stellar clusters: estimate for merger rate of binary black holes
  originating from globular clusters}
\author[A. Tanikawa]{A. Tanikawa$^{1,2,3}$\thanks{E-mail:
    ataru.tanikawa@riken.jp}\\ $^{1}$ RIKEN Advanced Institute for
  Computational Science 7--1--26, Minatojima-minami-machi, Chuo-ku,
  Kobe, Hyogo, 650--0047, Japan \\ $^{2}$School of Computer Science
  and Engineering, University of Aizu, Tsuruga Ikki-machi
  Aizu-Wakamatsu, Fukushima, 965-8580, Japan \\ $^{3}$Center for
  Computational Sciences, University of Tsukuba, 1--1--1, Tennodai,
  Tsukuba, Ibaraki 305--8577, Japan}
\begin{document}

\date{Accepted 1988 December 15. Received 1988 December 14; in
  original form 1988 October 11}

\pagerange{\pageref{firstpage}--\pageref{lastpage}} \pubyear{2002}

\maketitle

\label{firstpage}

\begin{abstract}
  We have performed $N$-body simulations of globular clusters (GCs) in
  order to estimate a detection rate of mergers of Binary stellar-mass
  Black Holes (BBHs) by means of gravitational wave (GW)
  observatories. For our estimate, we have only considered mergers of
  BBHs which escape from GCs (BBH escapers). BBH escapers merge more
  quickly than BBHs inside GCs because of their small semi-major
  axes. $N$-body simulation can not deal with a GC with the number of
  stars $N \sim 10^6$ due to its high calculation cost. We have
  simulated dynamical evolution of small-$N$ clusters ($10^4 \lesssim
  N \lesssim 10^5$), and have extrapolated our simulation results to
  large-$N$ clusters. From our simulation results, we have found the
  following dependence of BBH properties on $N$. BBHs escape from a
  cluster at each two-body relaxation time at a rate proportional to
  $N$. Semi-major axes of BBH escapers are inversely proportional to
  $N$, if initial mass densities of clusters are
  fixed. Eccentricities, primary masses, and mass ratios of BBH
  escapers are independent of $N$. Using this dependence of BBH
  properties, we have artificially generated a population of BBH
  escapers from a GC with $N \sim 10^6$, and have estimated a
  detection rate of mergers of BBH escapers by next-generation GW
  observatories. We have assumed that all the GCs are formed $10$ or
  $12$~Gyrs ago with their initial numbers of stars $N_{\rm i}= 5
  \times 10^5$ -- $2 \times 10^6$ and their initial stellar mass
  densities inside their half-mass radii $\rho_{\rm h,i}=6 \times 10^3$
  -- $10^6M_{\odot}\mbox{pc}^{-3}$. Then, the detection rate of BBH
  escapers is $0.5$ -- $20$~yr$^{-1}$ for a BH retention fraction
  $R_{\rm BH}=0.5$. A few BBH escapers are components of hierarchical
  triple systems, although we do not consider secular perturbation on
  such BBH escapers for our estimate. Our simulations have shown that
  BHs are still inside some of GCs at the present day. These BHs may
  marginally contribute to BBH detection.
\end{abstract}

\begin{keywords}
globular clusters: general -- binaries: close -- gravitational waves
-- stellar dynamics -- methods: numerical
\end{keywords}

\section{Introduction}
\label{sec:introduction}

Gravitational wave (GW) observation is greatly expected to open up new
fields in physics. Several ground-based GW observatories, such as
initial LIGO \citep{Abbott09}, Virgo \citep{Acernese08}, GEO600
\citep{Luck06}, and TAMA300 \citep{Takahashi04}, have operated, and
next-generation observatories, such as advanced LIGO \citep{Harry10},
advanced Virgo \citep{Accadia11}, and KAGRA \citep{Kuroda10}, will
start to run in the next several years. For these observatories,
mergers of compact-object binaries are one of the most promising GW
sources, where the compact-object binaries consist of stellar-mass
black holes (BHs) and neutron stars (NSs). Especially, binaries with
two BHs (hereafter, Binary BHs: BBHs) have been not yet found, since
they emit little electromagnetic wave. Detection of BBHs will have
invaluable impacts on astronomy and astrophysics.

For design of these observatories, it is important to make predictions
of a merger rate of BBHs as well as their chirp mass and mass ratio
distributions. Once GWs from BBH mergers are detected, these
predictions can constrain formation and evolution of BH progenitors,
i.e. massive stars. The evolution of the massive stars is, for
example, stellar wind and supernova explosion.  Furthermore, these
predictions should make clear whether GW detection can identify a
dominant BBH formation channel \citep[e.g.]{Sadowski08}
(S08). Theoretically, BBHs are thought to be formed through two
channels. One channel is BBH formation from a primordial binary
through stellar and binary evolution. In this channel, BBHs are formed
dominantly on galactic fields. The other is dynamical formation
through a dissipative two-body encounter, three-body encounter among
three single stars, binary-single encounter, and so on. The dynamical
formation channel happens only in dense stellar clusters, such as
globular clusters (GCs), young massive stellar clusters, and nuclear
stellar clusters.

So far, many efforts have been devoted to estimate detection rates of
BBHs formed in both the channels. \cite{Belczynski07} have estimated
that a detection rate of BBHs formed through the former channel by
means of advanced LIGO is $2$~yr$^{-1}$, although they has reported
the detection rate of $ \sim 10^4$~yr$^{-1}$ \citep{Belczynski12}. On
the other hand, \cite{PortegiesZwart00} (PZM00) have predicted that
BBHs originating from GCs merge frequently, and that their mergers
greatly contribute to GW detection. Following PZM00, various studies
estimate detection rates of BBH mergers. Their detection rates are not
necessarily consistent with each other: $1$ -- $10$~yr$^{-1}$
\citep{OLeary06} (O06), $25$ -- $3000$~yr$^{-1}$ (S08), and $1$ --
$100$~yr$^{-1}$ \citep{Downing10,Downing11} (D10 and D11,
respectively). \cite{Banerjee10} (B10) have reported that the
next-generation GW observatories detect mergers of BBHs formed in
young massive stellar clusters, rather than in GCs. \cite{OLeary09}
have proposed that mergers of BBHs formed in nuclear stellar clusters
are significant. These BBHs are detected by means of advanced LIGO at
a rate of $1$ -- $10$~yr$^{-1}$. Although the former channel seems to
form detectable BBHs more efficiently than the latter channel, an
estimate for a BBH detection rate is sensitive to details of
underlying models \citep{Dominik12}. It is worth while to investigate
a detection rate of mergers of BBHs formed through the latter channel
because of the simpleness of this channel. This channel involves only
Newtonian gravity (or low-order post-Newtonian gravity) on the very
moment of BBH formation.

In this paper, we especially focus on BBHs formed in GCs.  In order to
deal with BBH formation, many previous studies approximate GC
dynamical evolution in various ways. O06 and S08 follow BH evolution
in a GC with a fixed stellar density, determine whether a given BH
interacts with other single and binary stars by Monte Carlo technique,
and integrate numerically their orbits in the interaction. D10 and D11
follow BH interactions in a Monte Carlo technique similar to the above
two studies, however treat numerically GC dynamical evolution by Monte
Carlo method
\citep{Henon71,Stodolkiewicz82,Stodolkiewicz84,Giersz98}. These
methods are very convenient to survey vast parameter space of GC
initial conditions and stellar evolution because of their low
calculation costs, although their calculation costs differ in
degree. However, BH dynamics affects stellar densities of GCs, and
BBHs may be formed through more complicating interactions than
three-body interactions among single stars, binary-single
interactions, and binary-binary interactions. For example, six single
stars involve binary formation, even if all stars have equal masses
\citep{Tanikawa12,Tanikawa12b}.

$N$-body simulation can solve BH and GC dynamical evolution
self-consistently. However, it is practically impossible to treat an
$N$-body model with $N \sim 10^6$, comparable to the number of stars
in a real GC, due to its large calculation cost. B10 have followed
dynamical evolution of stellar clusters with $N \sim 10^5$, and have
estimated a BBH merger rate in young massive stellar clusters, not in
GCs. PZM00 have performed $N$-body simulations of stellar clusters
with several thousand stars, and have extrapolated a BBH merger rate
in a GC from their small-$N$ results. However, such an extrapolation
from small $N$ to large $N$ may involve difficult problems.

In this paper, we focus on, and verify $N$-dependence of BH dynamics
in a GC by means of $N$-body simulation. We investigate BBHs which
escape from GCs (hereafter, BBH escapers) in detail. BBH escapers
merge more quickly than BBHs inside GCs, since the former has smaller
semi-major axes than the latter. Using our simulation results, we
estimate a BBH detection rate, and predict detected BBH properties ,
such as distributions of chirp masses and mass ratios of BBH
escapers. We also investigate effects of a BH retention fraction.

This paper is organised as follows. In section \ref{sec:method}, we
describe our $N$-body simulation method. In section \ref{sec:results},
we show our simulation results. In section \ref{sec:estimate}, we
estimate BBH detection rates by means of GW observatories, based on
our simulation results shown in section \ref{sec:results}. In section
\ref{sec:discussion}, we discuss our estimate. Finally, we summarise
this paper in section \ref{sec:summary}.

\section{Method}
\label{sec:method}

\begin{table*}
  \caption{Summary of simulation models. Units of $r_{\rm v,i}$ and
    $\rho_{\rm h,i}$ are pc and $M_{\odot}$pc$^{-3}$, respectively.}
  \label{tab:model}
  \begin{center}
    \begin{tabular}{rrrrrccccc}
      \hline
      $N_{\rm i}$ &
      $r_{\rm v,i}$ &
      $\rho_{\rm h,i}$ &
      $R_{\rm BH}$ &
      $N_{\rm run}$ &
      $N_{\rm BH,tot,i}$ & $N_{\rm BH,i}$
      & $N_{\rm BBH,tot,esc}$ & $N_{\rm BBH,esc}$ & $N_{\rm CE-BBH,tot,esc}$ \\
      \hline
      $8k$   & $3.63$ &   $25$ & $1.0$ & $16$ & $197$ &  $7$ -- $14$ & $29$ &  $0$ --  $3$ & $0$ \\
      $16k$  & $4.57$ &   $25$ & $1.0$ &  $8$ & $188$ & $19$ -- $26$ & $24$ &  $3$ --  $3$ & $0$ \\
      $32k$  & $5.76$ &   $25$ & $1.0$ &  $4$ & $181$ & $42$ -- $48$ & $23$ &  $5$ --  $7$ & $0$ \\
      $64k$  & $7.26$ &   $25$ & $1.0$ &  $2$ & $187$ & $92$ -- $93$ & $21$ & $10$ -- $11$ & $0$ \\
      $128k$ & $9.15$ &   $25$ & $1.0$ &  $1$ & $178$ &      --      & $23$ &      --      & $0$ \\
      $8k$   & $1.44$ &  $400$ & $1.0$ & $16$ & $190$ &  $7$ -- $17$ & $32$ &  $0$ --  $4$ & $1$ \\
      $16k$  & $1.81$ &  $400$ & $1.0$ &  $8$ & $183$ & $19$ -- $26$ & $34$ &  $3$ --  $6$ & $1$ \\
      $32k$  & $2.28$ &  $400$ & $1.0$ &  $4$ & $180$ & $42$ -- $47$ & $28$ &  $5$ --  $8$ & $0$ \\
      $64k$  & $2.89$ &  $400$ & $1.0$ &  $2$ & $187$ & $93$ -- $94$ & $26$ & $11$ -- $15$ & $0$ \\
      $8k$   & $0.57$ & $6400$ & $1.0$ & $16$ & $142$ &  $4$ -- $15$ & $25$ &  $0$ --  $4$ & $7$ \\
      $16k$  & $0.72$ & $6400$ & $1.0$ &  $8$ & $145$ & $17$ -- $23$ & $22$ &  $1$ --  $5$ & $6$ \\
      $32k$  & $0.91$ & $6400$ & $1.0$ &  $4$ & $164$ & $38$ -- $45$ & $26$ &  $6$ --  $8$ & $4$ \\
      $64k$  & $1.14$ & $6400$ & $1.0$ &  $2$ & $178$ & $87$ -- $91$ & $28$ & $13$ -- $15$ & $0$ \\
      $64k$  & $7.26$ &   $25$ & $0.5$ &  $4$ & $180$ & $41$ -- $55$ & $25$ &  $5$ --  $9$ & $0$ \\
      $64k$  & $2.89$ &  $400$ & $0.5$ &  $4$ & $182$ & $41$ -- $56$ & $28$ &  $6$ --  $8$ & $1$ \\
      $64k$  & $1.14$ & $6400$ & $0.5$ &  $4$ & $179$ & $41$ -- $53$ & $25$ &  $4$ --  $8$ & $2$ \\
      $64k$  & $7.26$ &   $25$ & $0.25$ & $8$ & $189$ & $14$ -- $34$ & $27$ &  $0$ --  $6$ & $0$ \\
      $64k$  & $2.89$ &  $400$ & $0.25$ & $8$ & $193$ & $16$ -- $34$ & $32$ &  $3$ --  $5$ & $0$ \\
      $64k$  & $1.14$ & $6400$ & $0.25$ & $8$ & $180$ & $16$ -- $32$ & $26$ &  $1$ --  $5$ & $2$ \\
      \hline
    \end{tabular}
  \end{center}
\end{table*}

We summarise our simulation models in Table~\ref{tab:model}. We
choose the initial number of cluster stars as $N_{\rm i}=8k$, $16k$,
$32k$, $64k$, and $128k$, where $1k=1024$. We adopt Kroupa's function
\citep{Kroupa01} for a stellar initial mass function, given by
\begin{equation}
  f(m) dm \propto \left\{
  \begin{array}{cc}
    m^{-1.3}dm & \mbox{($m < 0.5 M_{\odot}$)} \\
    m^{-2.3}dm & \mbox{($m > 0.5 M_{\odot}$)} \\
  \end{array}
  \right.. \label{eq:kroupa01}
\end{equation}
We set maximum and minimum stellar masses to $50M_{\odot}$ and
$0.1M_{\odot}$, respectively. We include only single stars, and no
primordial binaries for simplicity. In this setup, the average stellar
mass of all the clusters is $0.61M_{\odot}$ at the initial time. These
stars are distributed according to King's model \citep{King66} with
dimensionless concentration parameter $W_0=7$. We introduce no
primordial mass segregation. Initial virial radii of these clusters,
$r_{\rm v,i}$, are shown in the second column of
Table~\ref{tab:model}, and are determined so as to set initial mass
densities inside half-mass radii, $\rho_{\rm h,i}$, to $25$, $400$,
and $6400M_{\odot}\mbox{pc}^{-3}$, regardless of $N_{\rm i}$. The
initial mass density, $\rho_{\rm h,i}$, is written in the third column
of Table~\ref{tab:model}. Hereafter, an unit of the stellar mass
density is $M_{\odot}\mbox{pc}^{-3}$, unless it is explicitly stated
otherwise. We adopt stellar metallicity as $Z=0.001$ which is one of
two peaks of the metallicity distribution of the galactic GC system
(S08) and GC systems in bright cluster galaxies \citep{Harris06}.

The clusters are embedded in an external tidal field of their parent
galaxy. The parent galaxy has a profile of flat circular velocity. The
circular velocity is set to $220$~km~s$^{-1}$. Our clusters move
around the parent galaxy on a circular orbit at the distance of
$8.5$~kpc from the galactic centre. In this external tidal field, the
clusters with $\rho_{\rm h,i}=25$ have King's cutoff radii equal to
Jacobi radii. Here, we define the King's cutoff radius as the distance
between the cluster centre and a point at which the stellar density
drops off into zero, and the Jacobi radius as the distance between the
cluster centre and the Lagrangian point (L1 or L2). This external
tidal field is typical for some of the galactic GCs.

\begin{figure}
 \begin{center}
   \includegraphics[scale=1.5]{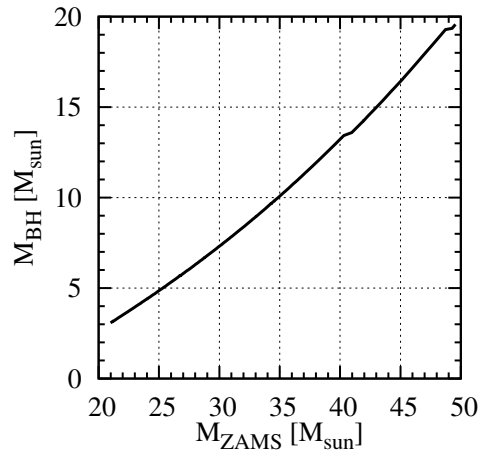}
 \end{center}
 \caption{Relation between ZAMS and BH masses. A symbol $M_{\rm sun}$
   in this figure is equivalent to $M_{\odot}$ in the main text, which
   is the case in the other figures.}
 \label{fig:zams-bh}
\end{figure}

As described in detail below, we actually use {\tt NBODY4}
\citep{Aarseth03} for our $N$-body simulations. A stellar evolution
model we adopt is attached to {\tt NBODY4}, although we change the
stellar evolution model in part. We choose stellar and binary
evolution models described in \cite{Hurley00} and \cite{Hurley01},
respectively. However, we choose mass losses of massive stars at their
supernova explosions from \cite{Eldridge04}. In the case of $Z=0.001$,
zero-age main-sequence (ZAMS) stars with $6.3$ -- $21M_{\odot}$ leave
NSs with $1.44M_{\odot}$. ZAMS stars with more than $21M_{\odot}$
become BHs, and the relation between ZAMS and BH masses is shown in
Fig.~\ref{fig:zams-bh}. Fractions of the numbers of NSs and BHs to the
total number of cluster stars are about $1$ and $0.2$ per cent,
respectively.

\begin{figure}
 \begin{center}
   \includegraphics[scale=1.5]{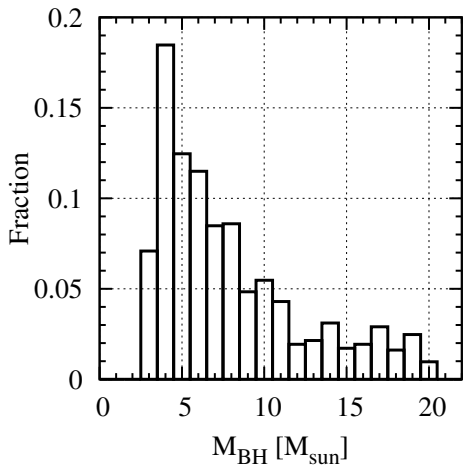}
 \end{center}
 \caption{BH mass function of our simulation models.}
 \label{fig:bhmassfunc}
\end{figure}

NSs and BHs should receive natal kicks due to asymmetric supernova
explosions. Since velocities of the natal kicks frequently exceeds
escape velocities of GCs, a significant fraction of NSs and BHs escape
from GCs. We model the natal kick as follows. A part of NSs and BHs
receive the natal kicks whose velocities exceed escape velocities of
clusters. These NSs and BHs are chosen by means of Monte Carlo
technique. NSs and BHs retained in the clusters receive no natal kick
at all. Note that the initial velocity distribution of BHs may be
different from that in reality due to such treatments for natal
kicks. The retention fraction is $R_{\rm NS}=0.1$ for the case of NSs
\citep{Pfahl02}. Our choice of a BH retention fraction, $R_{\rm BH}$,
is shown in the fourth column of Table~\ref{tab:model}, which is
consistent with $40$ -- $70$ per cent
\citep{Belczynski06,PortegiesZwart07}. The retention fractions of NSs
and BHs are assumed to be independent of ZAMS and remnant masses for
simplicity. Finally, we draw the BH mass function of our simulation
models in Fig.~\ref{fig:bhmassfunc}. The BH mass function is
independent of $R_{\rm BH}$.

Using different random seeds, we perform several runs for each
simulation model. The number of runs ($N_{\rm run}$) is determined,
such that the total number of BHs without natal kick in each
simulation model, $N_{\rm BH,tot,i}$, is the same among all the
simulation models. Then, $N_{\rm run}$ is inversely proportional to
$N_{\rm i}$ and $R_{\rm BH}$. The number of runs is shown in the fifth
column of Table~\ref{tab:model}. At each simulation model, $N_{\rm
  BH,tot,i}$ is about $140$ -- $200$, described in the sixth column of
Table~\ref{tab:model}. We also give the maximum and minimum numbers of
BHs without natal kick among runs at each simulation model, $N_{\rm
  BH,i}$, in the seventh column of Table~\ref{tab:model}.

We use the {\tt NBODY4} code for our $N$-body simulations. The {\tt
  NBODY4} code adopts fourth-order Hermite scheme with individual
timestep, and treat close encounters between two stars by KS
regularization \citep{Kustaanheimo65}, and those among more than two
stars by chain regularization \citep{Mikkola90,Mikkola93}.

We accelerate calculation of gravitational forces in the {\tt NBODY4}
code by using Graphic Processing Units (GPUs) with the {\tt Yebisu}
code \citep{Nitadori09}. Since the {\tt Yebisu} code does not
originally deal with no softened gravitational potential, we modify
the code so that it supports no softening force shape. We also
accelerate a search for neighbour particles around a given particle by
using SIMD instructions, here Advanced Vector eXtensions (AVX), with
Phantom-GRAPE for a collisional version \citep{Tanikawa12a}. We
parallelise the calculation of gravitational forces with Message
Passing Interface (MPI), such that we divide particles exerting
gravitational forces on a given particle into MPI processes, which is
so-called $j$-parallel algorithm. We search for neighbour particles in
parallel in the same way as the calculation of gravitational forces.

All of our runs are performed on ``HA-PACS'', which is a supercomputer
system at Center for Computational Sciences in University of
Tsukuba. One node of HA-PACS is configured as two Intel E5 CPUs and
four Nvidia M2090 GPUs. The nodes are connected in a fat-tree network
with Infiniband QDRx2. We use one node for simulation models with
$N_{\rm i}=8k$, $16k$, and $32k$, two nodes for simulation models with
$N_{\rm i}=64k$, and four nodes for simulation models with $N_{\rm
  i}=128k$. Simulation models with $N_{\rm i}=64k$ and $\rho_{\rm
  h,i}=6400$ are the most time-consuming. The wall-clock time for each
simulation is about $300$ hours.

Finally, we mention several units used in the following
sections. First, we often use a thermodynamical time, $\tau$, given by
\begin{equation}
  \tau = \int_0^t \frac{dt'}{t_{\rm rh}}, \label{eq:tau}
\end{equation}
where $t$ is a physical time, and $t_{\rm rh}$ is an instantaneous
half-mass relaxation time at $t$. The half-mass relaxation time is
defined as
\begin{equation}
  t_{\rm rh} = 0.0477 \frac{N}{\left( G\rho_{\rm h} \right)^{1/2}
    \log(0.4N)}, \label{eq:relaxationtime}
\end{equation}
where $G$ is the gravitational constant, $N$ is the number of stars of
a cluster at $t$, and $\rho_{\rm h}$ is a mass density inside a
half-mass radius of the cluster at $t$ \citep{Spitzer87}. The
thermodynamical time $\tau$ can be called as ``the elapsed number of
actual half-mass relaxation time'', which is similar to ``the elapsed
number of actual central relaxation time described at eq.~(4) of
\cite{Cohn89} \citep[see also]{Takahashi96}. Next, we describe units
related to binaries. We introduce an energy unit, $kT_{\rm i}$, where
$3/2kT_{\rm i}$ is the average kinetic energy of cluster stars at the
initial time. Then, we can write $1kT_{\rm i}$ as
\begin{equation}
  1kT_{\rm i} = \frac{1}{6N_{\rm i}} \frac{GM_{\rm i}^2}{r_{\rm v,i}}.
  \label{eq:ebkt}
\end{equation}
Furthermore, we introduce $a_{1kT_{\rm i}}$ as a length unit. A binary
has its semi-major axis $a_{1kT_{\rm i}}$, when it has its binding
energy $1kT_{\rm i}$ and component masses, both of which are the
average stellar mass at the initial time. The semi-major axis
$a_{1kT_{\rm i}}$ is given by
\begin{equation}
  a_{1kT_{\rm i}} = \frac{3}{N_{\rm i}} r_{\rm v,i}.
  \label{eq:abkt}
\end{equation}

\section{Simulation results}
\label{sec:results}

\subsection{Thermodynamical evolution}
\label{sec:thermo}

In this section, we investigate a relation between a physical time $t$
and a thermodynamical time $\tau$ given by equation~(\ref{eq:tau}). A
pure $N$-body system is scale-free, and is evolved only by two-body
relaxation, which results from gravitational interactions between
stars. Our simulation models include stellar evolution, and are not
pure $N$-body systems. Nevertheless, our simulation models have
aspects of pure $N$-body systems. In fact, BHs and BBHs in clusters
with different $N_{\rm i}$ and $\rho_{\rm h,i}$ evolve similarly in
terms of $\tau$, as shown in section~\ref{sec:BHandBBH}. The
thermodynamical time $\tau$ is an indicator for dynamical states of
clusters with different $N_{\rm i}$ and $\rho_{\rm h,i}$. Converting
$\tau$ to $t$, we can know states of BHs and BBHs of a cluster at a
given physical time, even if the cluster have $N_{\rm i}$ and
$\rho_{\rm h,i}$ which we do not adopt in our simulations.

\begin{figure*}
 \begin{center}
   \includegraphics[scale=1.5]{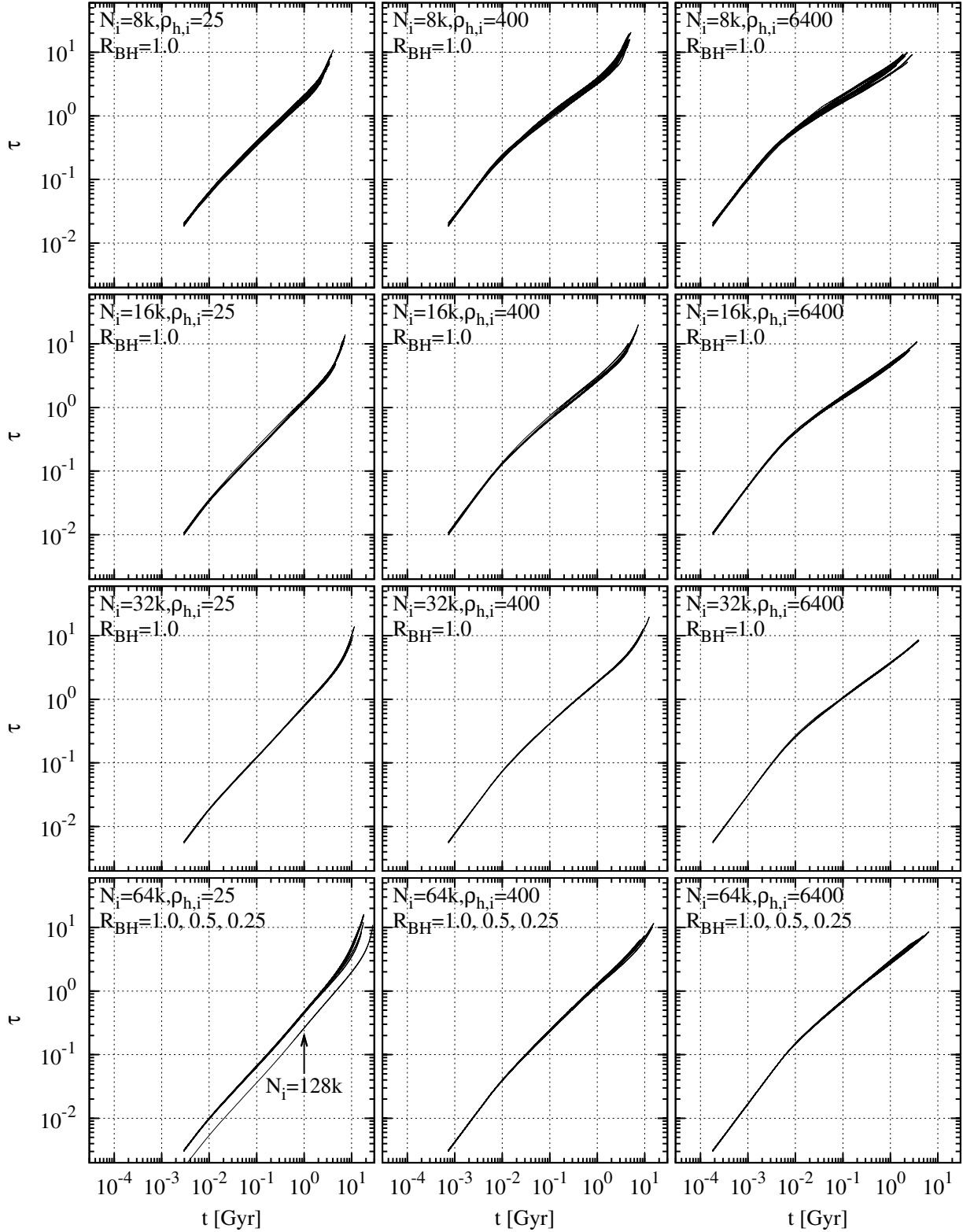}
 \end{center}
 \caption{Relation between a physical time, $t$, and a thermodynamical
   time, $\tau$, in all the runs. Simulation models are indicated in
   each panel, except a simulation model with $N_{\rm i}=128k$,
   $\rho_{\rm h,i}=25$ and $R_{\rm BH}=1.0$, which is pointed to by
   the arrow. The panels of $N_{\rm i}=64k$ include all the simulation
   models with $R_{\rm BH}=1.0$, $0.5$, and $0.25$, and the other
   panels include only simulation models with $R_{\rm BH}=1.0$.}
 \label{fig:tau_time}
\end{figure*}

Fig.~\ref{fig:tau_time} shows relations between $t$ and $\tau$ in all
the runs. These relations are almost the same among all the runs in
each simulation model. Clusters with different random seeds evolve
similarly to each other in terms of thermodynamics.  Relations between
$t$ and $\tau$ are also similar among simulation models with different
$R_{\rm BH}$ (see the bottom panels of Fig.~\ref{fig:tau_time}). Even
if $R_{\rm BH}$ is different, cluster evolution is thermodynamically
similar. In other words, $N$ and $\rho_{\rm h}$ evolve similarly. This
is because the total number and mass of BHs are much smaller than
those of cluster stars.

\subsection{BH and BBH evolution}
\label{sec:BHandBBH}

\begin{figure*}
 \begin{center}
   \includegraphics[scale=1.4]{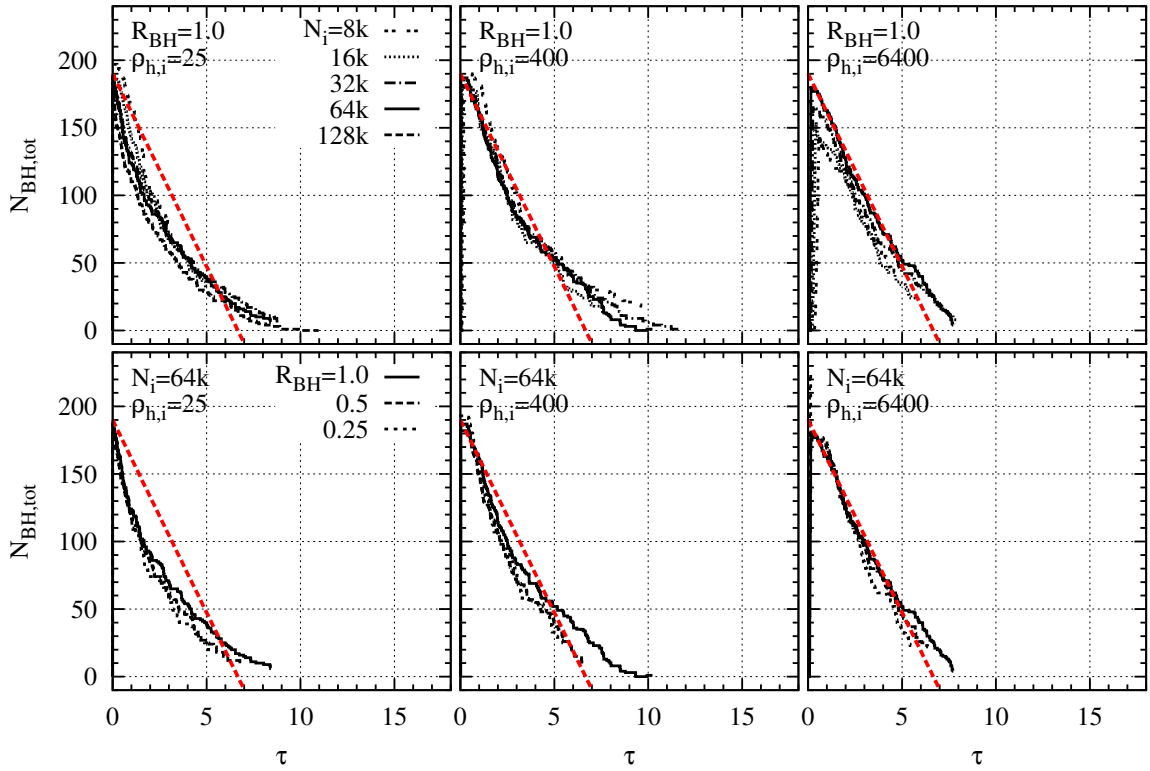}
 \end{center}
 \caption{Time evolution of the numbers of BHs in the clusters. The
   number is the summation of the numbers of BHs of all the runs in
   each simulation model, not the number of BHs in each run. The
   horizontal axis is a thermodynamical time, $\tau$, defined as
   equation~(\ref{eq:tau}). Curve with the same line types indicate
   the same $N_{\rm i}$ among the top panels, and the same $R_{\rm
     BH}$ among the bottom panels. Red dashed line in each panel shows
   $N_{\rm BH,tot} = N_{\rm BH,tot,i}(1-0.15\tau)$, where $N_{\rm
     BH,tot,i}=190$.}
 \label{fig:nsbh_tau}
\end{figure*}

In this section, we show BH and BBH evolution in terms of
$\tau$. Fig.~\ref{fig:nsbh_tau} shows the time evolution of the
numbers of BHs in clusters, $N_{\rm BH,tot}$. The number $N_{\rm
  BH,tot}$ is the summation of the numbers of BHs of all the runs in
each simulation model, not the number of BHs in each run. First, we
focus on simulation models with the same $\rho_{\rm h,i}$ and $R_{\rm
  BH}$, and different $N_{\rm i}$ (see each of the top panel in
Fig.~\ref{fig:nsbh_tau}). We can see that $N_{\rm BH,tot}$ decreases
in a similar manner. This means that the number of BHs which escape
from each cluster at each thermodynamical time is proportional to
$N_{\rm i}$. Next, we investigate simulation models with the same
$\rho_{\rm h,i}$ and $N_{\rm i}$, and different $R_{\rm BH}$ (see each
of the bottom panel in Fig.~\ref{fig:nsbh_tau}). The number $N_{\rm
  BH,tot}$ also decreases in a similar manner among different $R_{\rm
  BH}$ models. The number of BHs which escape from a cluster at each
thermodynamical time is proportional to $R_{\rm BH}$.

We pay attention to dependence of BH evolution on $\rho_{\rm h,i}$
(see all the top panels of Fig.~\ref{fig:nsbh_tau}). We draw the same
red lines in all the panels of Fig.~\ref{fig:nsbh_tau}, which fit to
models with $\rho_{\rm h,i}=6400$ and $R_{\rm BH}=1.0$. The red line
is expressed as $N_{\rm BH,tot} = N_{\rm BH,tot,i}(1-0.15\tau)$, where
$N_{\rm BH,tot,i}=190$, which is consistent with our simulation models
(see the sixth column of Table~\ref{tab:model}). Note that this is
just a simple fitting. We do not claim that there is some physics
behind it. The red line is in good agreement with BH evolution of
models with $\rho_{\rm h,i}=400$ and $R_{\rm BH}=1.0$ before
$\tau=6$. At first glance, the red line does not seem to be at all in
agreement with BH evolution of models with $\rho_{\rm h,i}=25$ and
$R_{\rm BH}=1.0$ all the time. However, decrease rates of $N_{\rm
  BH,tot}$ at $\tau=0$ and $6$ in these models are, respectively,
larger and smaller than that of the red line only by a factor of about
$2$. The decrease rates of $N_{\rm BH,tot}$ are different at most by a
factor of $2$, despite that $\rho_{\rm h,i}$ ranges from $25$ to
$6400$. We can regard that BHs evolve almost independently of
$\rho_{\rm h,i}$. In summary, BHs escape from a cluster at a rate
proportional to $N_{\rm i}$ and $R_{\rm BH}$, and at a rate
independent of $\rho_{\rm h,i}$ in the unit of $\tau$. Therefore, we
can give the time evolution of the number of BHs in a cluster as
follows:
\begin{equation}
  N_{\rm BH}(\tau) = N_{\rm BH,i} \left(1 - 0.15
  \tau\right). \label{eq:fit_sbh}
\end{equation}

The eighth and ninth columns of Table~\ref{tab:model} show the total
number of BBH escapers in each simulation model, and the minimum and
maximum numbers of BBH escapers among all the runs of each simulation
model, respectively. The total number of BBH escapers, $N_{\rm
  BBH,tot,esc}$, is not dominated by only one of runs in any
simulation models. The difference of the number of BBH escapers among
runs in each simulation model is small. This difference comes from the
difference of the number of BHs without the natal kick, and from
statistical fluctuation, not from the difference of global evolution
among runs.

\begin{figure*}
 \begin{center}
   \includegraphics[scale=1.4]{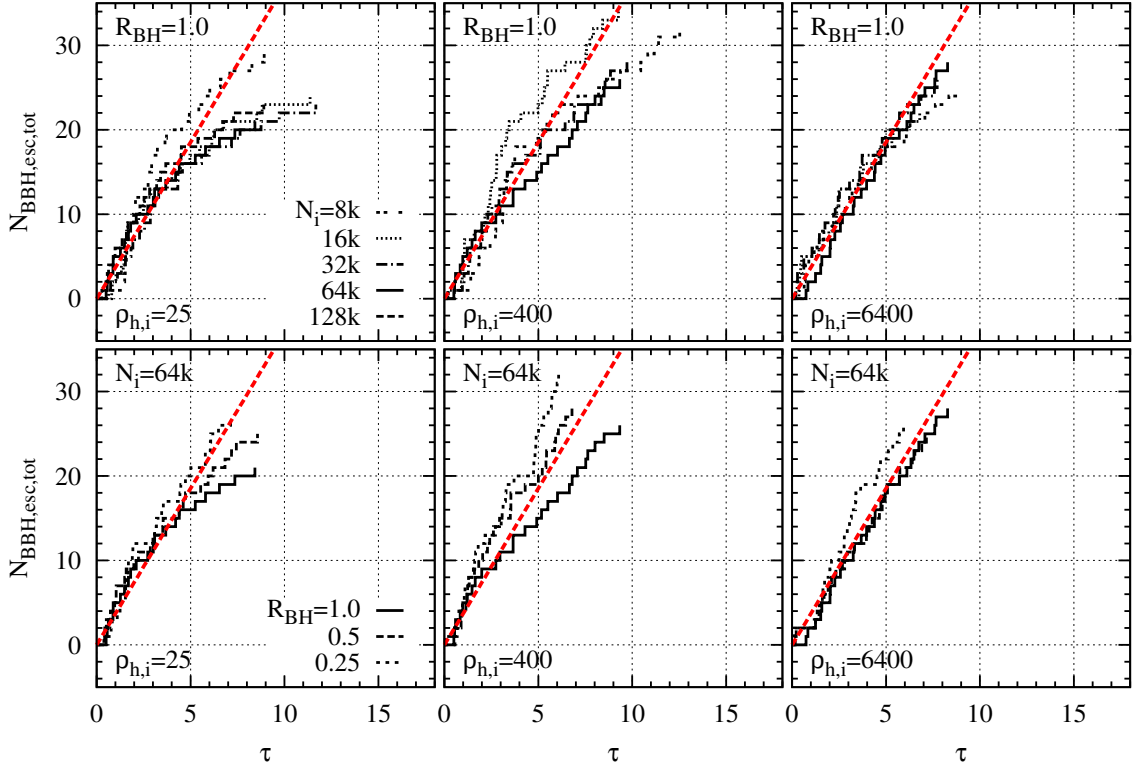}
 \end{center}
 \caption{Time evolution of the cumulative number of BBH escapers. The
   cumulative number is the summation of the cumulative numbers of BBH
   escapers of all the runs in each simulation model, not the
   cumulative number of BBH escapers in each run. The horizontal axes
   are the same as Fig.~\ref{fig:nsbh_tau}. Curves with the same line
   types indicate the same $N_{\rm i}$ among the top panels, and the
   same $R_{\rm BH}$ among the bottom panels. Red dashed line in each
   panel shows $N_{\rm BBH,esc,tot} = 0.020N_{\rm BH,tot,i}\tau$,
   where $N_{\rm BH,tot,i}=190$.}
 \label{fig:nbbh_tau}
\end{figure*}

Fig.~\ref{fig:nbbh_tau} shows the time evolution of the cumulative
number of BBH escapers, $N_{\rm BBH,esc,tot}$. The number $N_{\rm
  BBH,esc,tot}$ is the summation of the numbers of BBH escapers of all
the runs in each simulation model, not the number of BBH escapers in
each run. As seen in each panel of Fig.~\ref{fig:nbbh_tau}, $N_{\rm
  BBH,esc,tot}$ increases in a similar way among models with the same
$\rho_{\rm h,i}$ and $R_{\rm BH}$, and different $N_{\rm i}$, and
among models with the same $\rho_{\rm h,i}$ and $N_{\rm i}$, and
different $R_{\rm BH}$. We fit a red line to $N_{\rm BBH,esc,tot}$ of
models with $\rho_{\rm h,i}=6400$ and $R_{\rm BH}=1.0$. The red line
is expressed as $N_{\rm BBH,esc,tot} = 0.020N_{\rm
  BH,tot,i}\tau$. Note that this is just a simple fitting. We do not
claim that there is some physics behind it. During $\tau=0$ -- $5$,
the red lines are in good agreement with $N_{\rm BBH,esc,tot}$ in all
the models. In summary, BBHs escape from a cluster in a rate
proportional to $N_{\rm i}$ and $R_{\rm BH}$, and at a rate
independently of $\rho_{\rm h,i}$, which is similar to BH
evolution. At least during $\tau=0$ -- $5$, we can give the time
evolution of the cumulative number of BBH escapers as follows:
\begin{equation}
  N_{\rm BBH,esc} (\tau) = 0.020 N_{\rm BH,i}
  \tau. \label{eq:fit_bbhesc}
\end{equation}

We use equations~(\ref{eq:fit_bbhesc}) only for an estimate of a BBH
detection rate in section \ref{sec:estimate}. The fitting is in good
agreement with our simulation results during $\tau=0$ -- $5$ in all
the models, and during $\tau=0$ -- $8$ in models with $\rho_{\rm
  h,i}=400$ and $6400$. We do not mind that the red lines are slightly
deviated from our simulation results in models with $\rho_{\rm
  h,i}=25$ after $\tau=5$. These clusters are evaporated strongly in
contrast to those in models with $\rho_{\rm h,i}=400$ and $6400$. In
section~\ref{sec:estimate}, we estimate the detection rate of BBH
escapers from clusters with $\rho_{\rm h,i}$ larger than $6400$, which
are evaporated less than clusters with $\rho_{\rm h,i}=6400$. The
number of BBH escapers from these clusters should evolve similarly to
clusters in models with $\rho_{\rm h,i}=400$ and $6400$.

We can explain why the numbers of BH and BBH escapers at each
thermodynamical time are proportional to $N_{\rm i}$ as follow. After
BHs and NSs are formed, stars do not evolve so active. Our simulation
models can be regarded as nearly pure $N$-body systems. Therefore, our
simulation models are scale-free, and evolved by two-body
relaxation. At each thermodynamical time, a constant fraction of the
total energy of a cluster flows out from the cluster through two-body
relaxation. If there is no energy source at the cluster core, the core
continues to shrink, and its density grows. At some point, binaries
are formed because of high density at the core. The binaries interact
with other single and binary stars in a superelastic manner. Such
interactions result in ejections of other stars and binaries
themselves. These ejections indirectly heat the cluster, since the
cluster loses binding energies of ejected stars. This indirect heating
is balanced with the energy outflowing through two-body relaxation, so
that the core stops shrinking. Therefore, the cluster loses a constant
fraction of its total mass through the ejections of single and binary
stars at each thermodynamical time. Since the mass lost by the cluster
contains a constant fraction of BHs, the number of BH escapers is
proportional to $N_{\rm i}$ at each thermodynamical time, which can be
true of BBH escapers.

\begin{figure*}
 \begin{center}
   \includegraphics[scale=1.4]{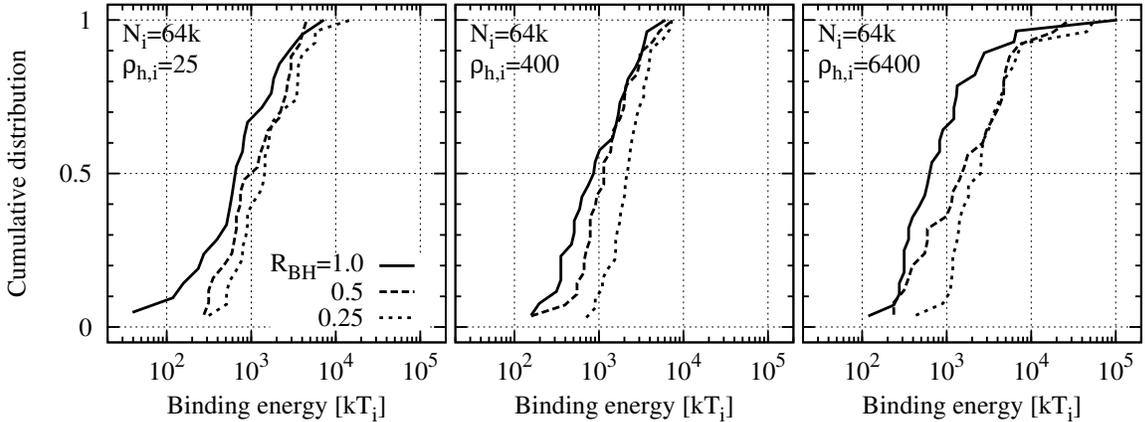}
 \end{center}
 \caption{Cumulative distribution of binding energies of BBH escapers
   in all the runs of each simulation model. The total BBH escapers in
   all the runs of each simulation model are accumulated, which is the
   case for all the cumulative distributions in this paper. The unit
   of binding energies is $kT_{\rm i}$, defined as
   equation~(\ref{eq:ebkt}).}
 \label{fig:cf_eb}
\end{figure*}

The numbers of BH and BBH escapers at each thermodynamical time are
proportional to $R_{\rm BH}$. This is because the mass lost by a
cluster contains a fraction of BHs, proportional to $R_{\rm BH}$.
Since clusters with different $R_{\rm BH}$ evolve similarly in terms
of $\tau$ (see the bottom panels of Fig.~\ref{fig:tau_time}), BBHs
should be formed similarly among these clusters. However, BBH escape
rates are proportional to $R_{\rm BH}$. Therefore, a given BBH stays
for a longer time in a cluster with smaller $R_{\rm BH}$. This may be
because there are fewer BHs to scatter with in these clusters
\citep{Mackey08}. This dependence on $R_{\rm BH}$ seems consistent
with simulation results of \cite{Breen13}, when we compare points with
the same sizes and symbols between filled and unfilled ones in
fig.~8. As a BBH stays for a longer time in a cluster, the BBH emits
larger energy, and its binding energy becomes larger. Consequently,
BBH escapers from clusters with smaller $R_{\rm BH}$ should have
larger binding energies. Actually, this is consistent with our
simulation results. Fig.~\ref{fig:cf_eb} shows the cumulative
distribution of binding energies of BBH escapers. In each panel, we
can see that BBH escapers in models with smaller $R_{\rm BH}$ have
larger binding energies.

\subsection{Properties of BBH escapers}
\label{sec:propBBH}

In this section, we investigate orbital elements and mass components
of BBHs, which determine a merging timescale through GW. We focus on
BBH escapers. The most hardest binaries in a cluster are the most
likely to be ejected due to recoil during an interaction. These
binaries have the smallest semi-major axes, if all the binaries have
BHs with the same masses. BBH escapers merge more quickly than BBHs
inside GCs.

\begin{figure*}
 \begin{center}
   \includegraphics[scale=1.4]{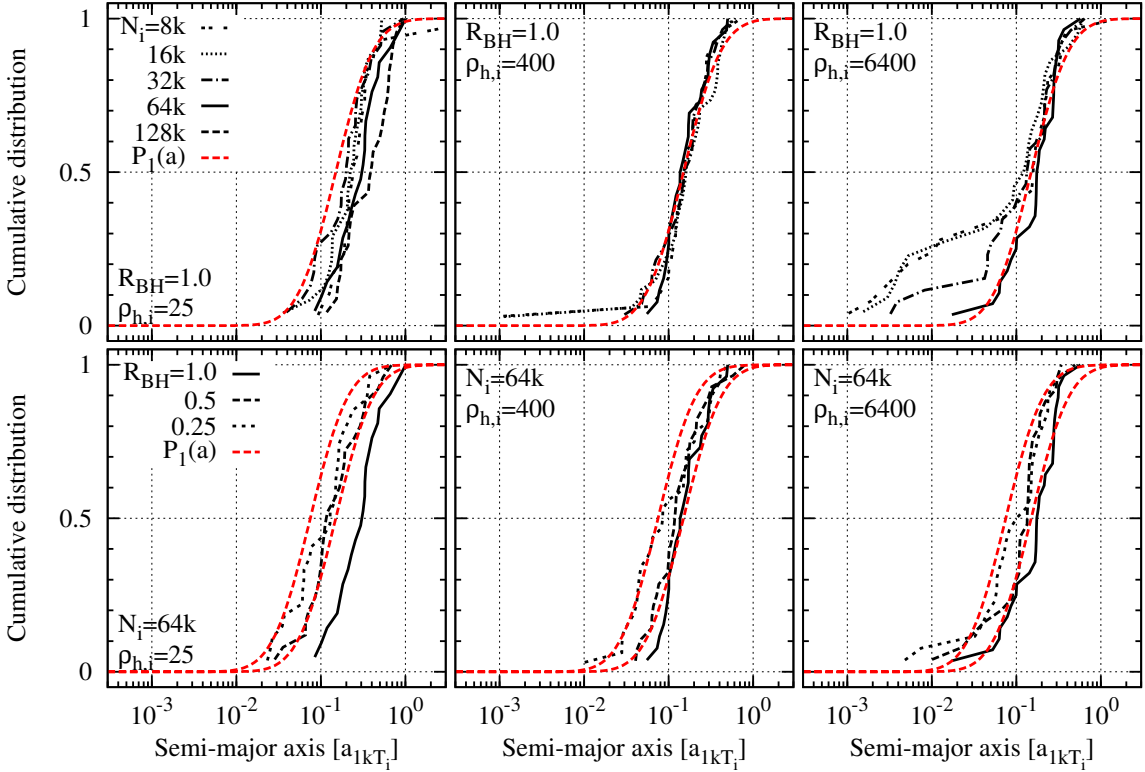}
 \end{center}
 \caption{Cumulative distribution of semi-major axes of BBH
   escapers. The unit of semi-major axes is $a_{\rm 1kT_{\rm i}}$,
   defined as equation~(\ref{eq:abkt}). Curves with the same line
   types indicate the same $N_{\rm i}$ among the top panels, and the
   same $R_{\rm BH}$ among the bottom panels. The formula of red
   curves is expressed as equation~(\ref{eq:fit_axis}). $R_{\rm
     BH}=1.0$ is adopted for the red curves in the top panels, and
   $R_{\rm BH}=0.25$ (left red curves) and $1.0$ (right red curves) in
   the bottom panels.}
 \label{fig:cf_axis}
\end{figure*}

We show the cumulative distribution of semi-major axes of BBH escapers
in Fig.~\ref{fig:cf_axis}. First, we take a look at their dependence
on $N_{\rm i}$ among each simulation model with the same $\rho_{\rm
  h,i}$ and $R_{\rm BH}$ (see each of the top panels of
Fig.~\ref{fig:cf_axis}). In simulation models with $\rho_{\rm h,i}=25$
and $400$, their cumulative distributions are independent of $N_{\rm
  i}$ in the unit of $a_{\rm 1kT_{\rm i}}$. In simulation models with
$\rho_{\rm h,i}=6400$, their cumulative distributions are quite
different. $20$ percentiles in these models become larger with $N_{\rm
  i}$ increasing: $4 \times 10^{-3} a_{\rm 1kT_{\rm i}}$ for $N_{\rm
  i}=8k$ and $16k$, $4 \times 10^{-2} a_{\rm 1kT_{\rm i}}$ for $N_{\rm
  i}=32k$, and $9 \times 10^{-2} a_{\rm 1kT_{\rm i}}$ for $N_{\rm
  i}=64k$. This is because smaller $N_{\rm i}$ models contain more
escapers of BBHs whose formation is involved by common envelop
evolution. We call these BBH escapers ``CE-BBH escapers''. In contrast
to CE-BBH escapers, we call escapers of BBHs formed only through an
$N$-body process ``NB-BBH escapers''.

\begin{figure*}
 \begin{center}
   \includegraphics[scale=1.2]{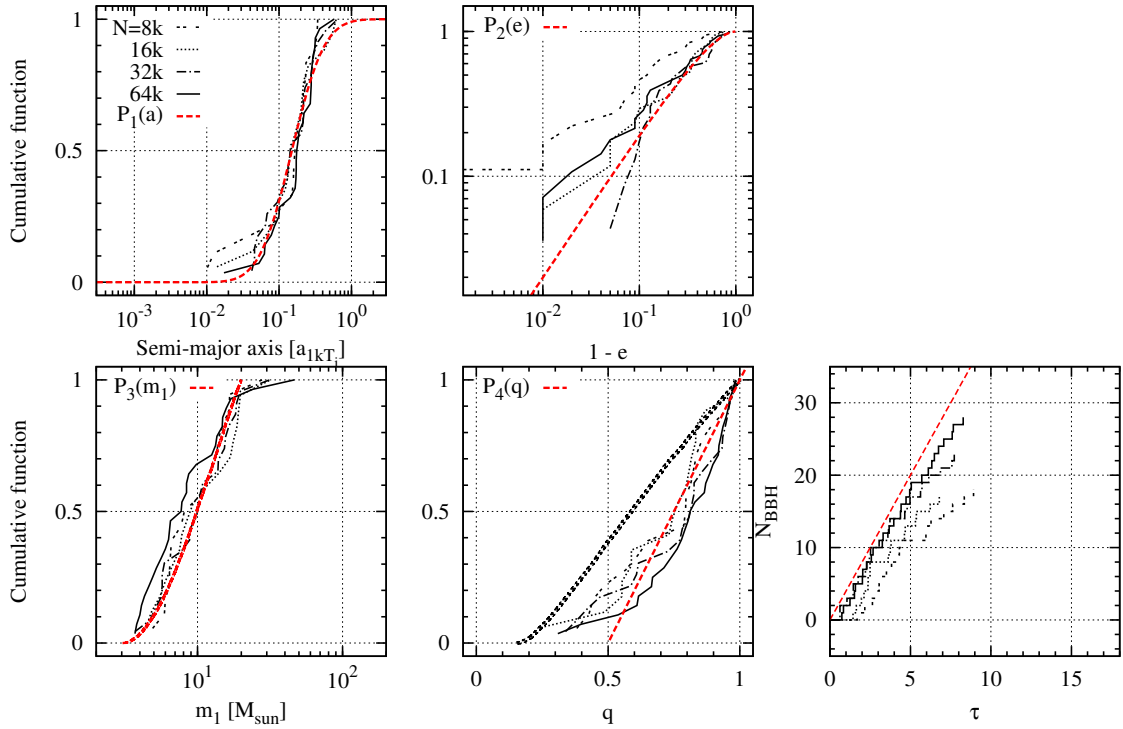}
 \end{center}
 \caption{Cumulative distribution of semi-major axes, eccentricities,
   primary masses, and mass ratios of NB-BBHs, and time evolution of
   the cumulative number of NB-BBHs in models with $\rho_{\rm
     h,i}=6400$ and $R_{\rm BH}=1.0$. The red dashed line in the
   bottom right panel is identical with those in
   Figure~\ref{fig:nbbh_tau}.}
 \label{fig:cf_cmev}
\end{figure*}

These CE-BBHs are formed as follows. Two BH progenitors (or a pair of
a BH and BH progenitor) form a binary through three-body
interactions. Either of the two evolves off the main sequence, and
common envelop evolution occurs. In some of these binaries, common
envelop evolution occurs again, when the other star evolves off the
main sequence. CE-BBH escapers tend to have smaller semi-major axes
than NB-BBH escapers. Common envelop evolution shrink a semi-major
axis of a binary, involving no kick on the binary, except BH natal
kick. Note that CE-BBHs receive no BH natal kick at a constant rate in
our natal kick model. On the other hand, an $N$-body process, such as
a three-body interaction, also shrinks a semi-major axis of a binary,
but gives dynamical recoil on the binary. Therefore, NB-BBHs are
ejected from the cluster before their semi-major axes become as small
as those of CE-BBHs. The top left panel of Fig.~\ref{fig:cf_cmev}
shows the cumulative distribution of semi-major axes of only NB-BBH
escapers. We can see that their semi-major axes are distributed
independently of $N_{\rm i}$.

The tenth column of Table~\ref{tab:model} indicates the total number
of CE-BBH escapers in all the runs of each simulation model. The
number of CE-BBHs decreases as $N_{\rm i}$ increases, and $\rho_{\rm
  h,i}$ decreases. A CE-BBH escaper has already been a binary before
at least one of its components becomes a BH. In order for CE-BBH
escapers to be formed, a cluster experiences core collapse before
massive stars evolve to BHs. Therefore, the number of CE-BBH escapers
decreases in a cluster with longer half-mass relaxation time,
i.e. that with larger $N_{\rm i}$ and smaller $\rho_{\rm h,i}$.

We focus on dependence of semi-major axes of BBH escapers on
$\rho_{\rm h,i}$ (see the top left and middle panels of
Fig.~\ref{fig:cf_axis}, and the top left panel of
Fig.~\ref{fig:cf_cmev}). We draw red curves in all these panels.
Formula of these red curves is given by
\begin{equation}
  P_1(a) = \int_0^a p_1(a') da', \label{eq:fit_cum_axis}
\end{equation}
where
\begin{align}
  p_1(a) = &\frac{1}{\sqrt{2\pi} \sigma (a/a_{\rm 1kT_{\rm i}})}
  \nonumber \\ &\times \exp \left\{ -\frac{1}{2\sigma^2} \left[ \log
    \left( \frac{a}{a_{1kT_{\rm i}}} R_{\rm BH}^{-1/2} \right) -
    \log\mu \right]^2 \right\}. \label{eq:fit_axis}
\end{align}
The function $p_1(a)$ is a log-normal distribution with $\sigma=0.81$
and $\mu=0.15$. The red curves fit to cumulative distributions of BBH
escapers in simulation models with $\rho_{\rm h,i}=400$ and $R_{\rm
  BH}=1.0$ (the top middle panel of Fig.~\ref{fig:cf_axis}), and of
NB-BBH escapers in simulation models with $\rho_{\rm h,i}=6400$ and
$R_{\rm BH}=1.0$ (the top left panel of Fig.~\ref{fig:cf_cmev}).

As seen in the bottom panels of Fig.~\ref{fig:cf_axis}, BBH escapers
in smaller $R_{\rm BH}$ models have smaller semi-major axes than those
in larger $R_{\rm BH}$ models. This is reflected by the fact that BBH
escapers in smaller $R_{\rm BH}$ models have larger binding energies
than those in larger $R_{\rm BH}$ models (see
Fig.~\ref{fig:cf_eb}). The red curves are in good agreement with the
cumulative distribution, when the semi-major axes of BBH escapers are
are proportional to $R_{\rm BH}^{-1/2}$, as indicated in
equation~(\ref{eq:fit_axis}). We note that
equation~(\ref{eq:fit_axis}) is just a simple fitting. We do not claim
that there is any physical background for this probability
distribution.

Note that the distribution of the semi-major axes in simulation models
with $\rho_{\rm h,i}=25$ is larger than the fitting formula. Clusters
with $\rho_{\rm h,i}=25$ strongly experience mass loss due to
evaporation, since external tidal fields are strong for these
models. Small escape velocities of these clusters results in ejections
of BBHs which are not so hard, i.e. have large semi-major axes. On the
other hand, clusters with $\rho_{\rm h,i}=400$ and $6400$ are not
evaporated so much. Owing to the fact that pure $N$-body system is
scale-free, the distributions of the semi-major axes of BBH escapers
are similar between clusters with $\rho_{\rm h,i}=400$ and $6400$, and
are fitted to equation~(\ref{eq:fit_cum_axis}). Eventually, we
estimate a detection rate of BBH escapers from clusters with
$\rho_{\rm h,i}$ larger than $6400$ in
section~\ref{sec:estimate}. These clusters are less evaporated than
the clusters with $\rho_{\rm h,i}=400$ and $6400$, since they are more
compact. BBH escapers from these clusters should have similar
distributions of semi-major axes to this fitting formula. This is the
reason why we ignore data of simulation models with $\rho_{\rm
  h,i}=25$ for the fitting formula.

So far, we describe semi-major axes of BBH escapers in the units of
$a_{\rm 1kT_{\rm i}}$. Converting the units of $a_{\rm 1kT_{\rm i}}$
to physical units, we can say that semi-major axes of BBH escapers are
inversely proportional to $N_{\rm i}$, and proportional to the initial
size of clusters, such as virial radii (see equation~(\ref{eq:abkt})).

\begin{figure*}
 \begin{center}
   \includegraphics[scale=1.4]{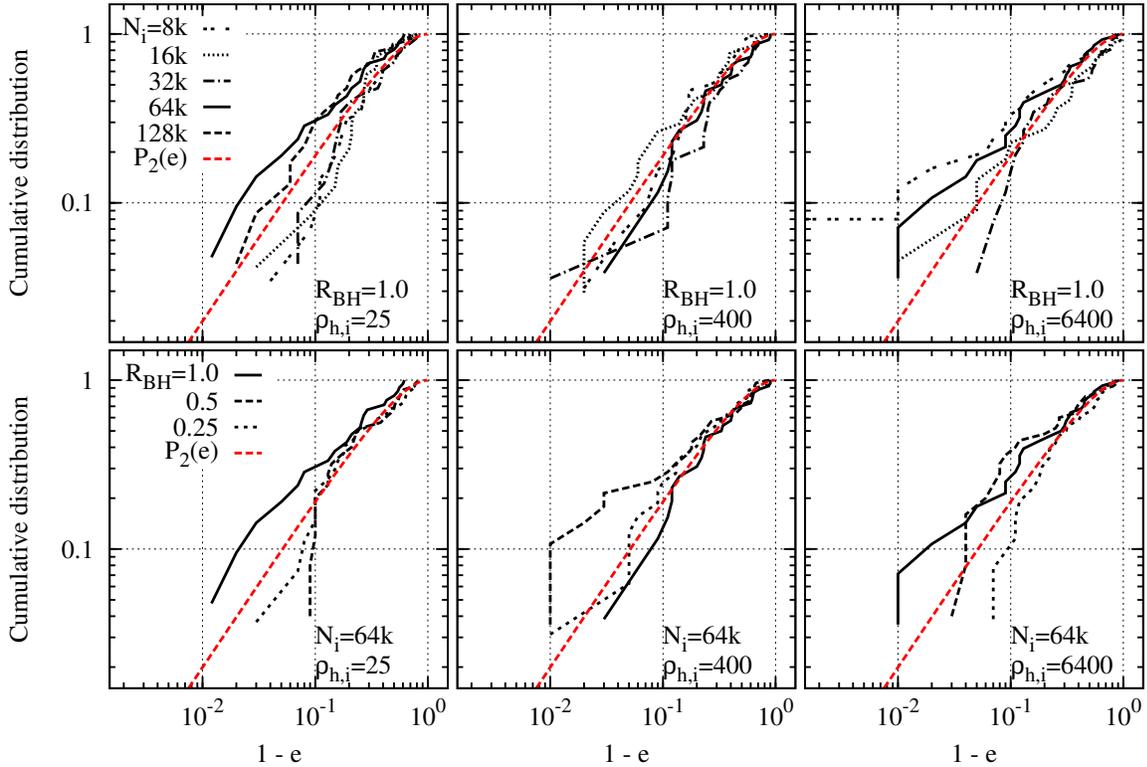}
 \end{center}
 \caption{Cumulative distribution of eccentricities, $e$, of BBH
   escapers. Curves with the same line types indicate the same $N_{\rm
     i}$ among the top panels, and the same $R_{\rm BH}$ among the
   bottom panels. Red dashed curves indicate the thermal distribution,
   expressed as equation~(\ref{eq:fit_ecc}).}
 \label{fig:cf_ecc}
\end{figure*}

We move on to the subject of eccentricities of BBH escapers. We see
cumulative distributions of eccentricities of BBH escapers in
Fig.~\ref{fig:cf_ecc}. The eccentricity distributions in all the
simulation models are consistent with the thermal distribution
\citep{Heggie75}, expressed as
\begin{equation}
  p_2(e) = 2e. \label{eq:fit_ecc}
\end{equation}
Formula of the red curves in Fig.~\ref{fig:cf_ecc} is expressed as
\begin{equation}
  P_2(e) = \int_0^e p_2(e') de'.
\end{equation}
Although the cumulative distributions are deviated from the function
$P_2(e)$, they are centred around it.

\begin{figure*}
 \begin{center}
   \includegraphics[scale=1.4]{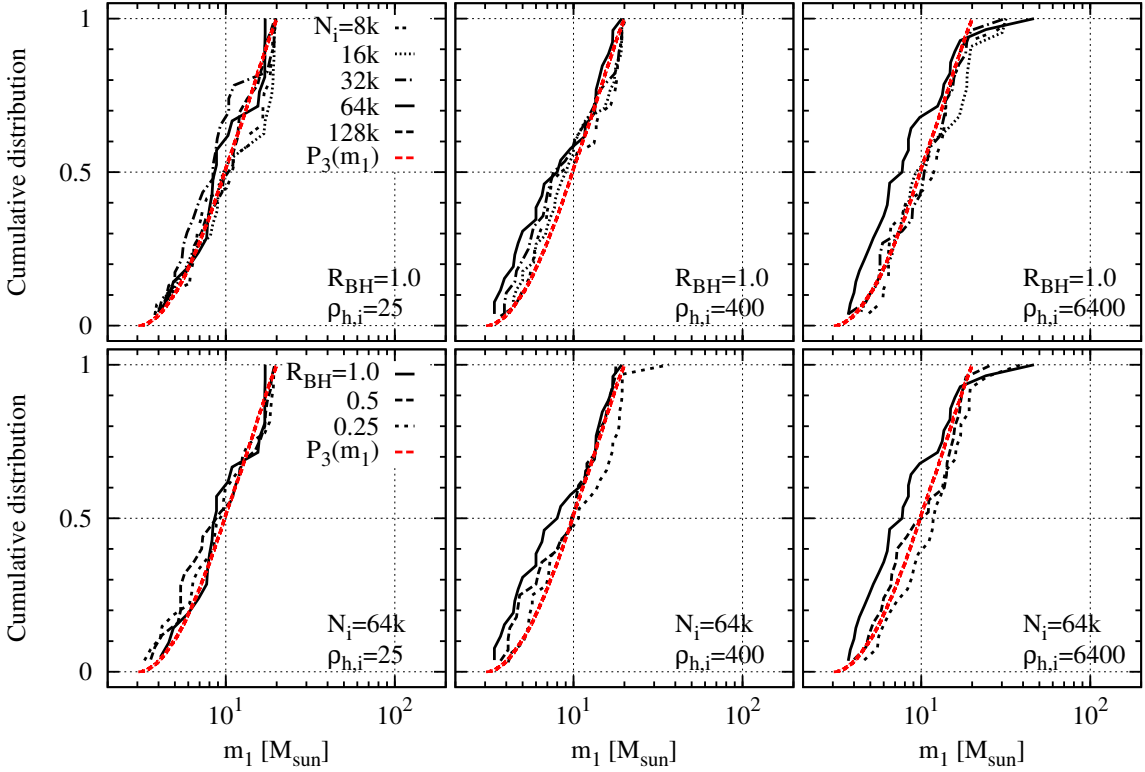}
 \end{center}
 \caption{Cumulative distribution of primary masses, $m_1$, of BBH
   escapers. Curves with the same line types indicate the same $N_{\rm
     i}$ among the top panels, and the same $R_{\rm BH}$ among the
   bottom panels. Red dashed curves show a fitting function for these
   cumulative distribution. The method to draw the fitting function is
   described in the main text.}
 \label{fig:cf_m1}
\end{figure*}

Next, we investigate cumulative distributions of primary masses of BBH
escapers, which is shown in Fig.~\ref{fig:cf_m1}. The cumulative
distributions are independent of $N_{\rm i}$, $\rho_{\rm h,i}$, and
$R_{\rm BH}$. We compare the distribution of primary masses of
artificially generated BBHs, where we define their distribution as
$p_3(m_1)$, and their cumulative distribution as $P_3(m_1)$, where
$m_1$ is the primary BH mass of a BBH escaper. We generate these BBHs
as follows. We realise two BBHs whose mass distribution is subject to
that shown in Fig.~\ref{fig:bhmassfunc}, and choose the more massive
BH as the primary BH. The function $P_3(m_1)$, drawn by red curves in
Fig.~\ref{fig:cf_m1}, is in good agreement with the cumulative
distribution of primary masses of BBH escapers.

\begin{figure*}
 \begin{center}
   \includegraphics[scale=1.4]{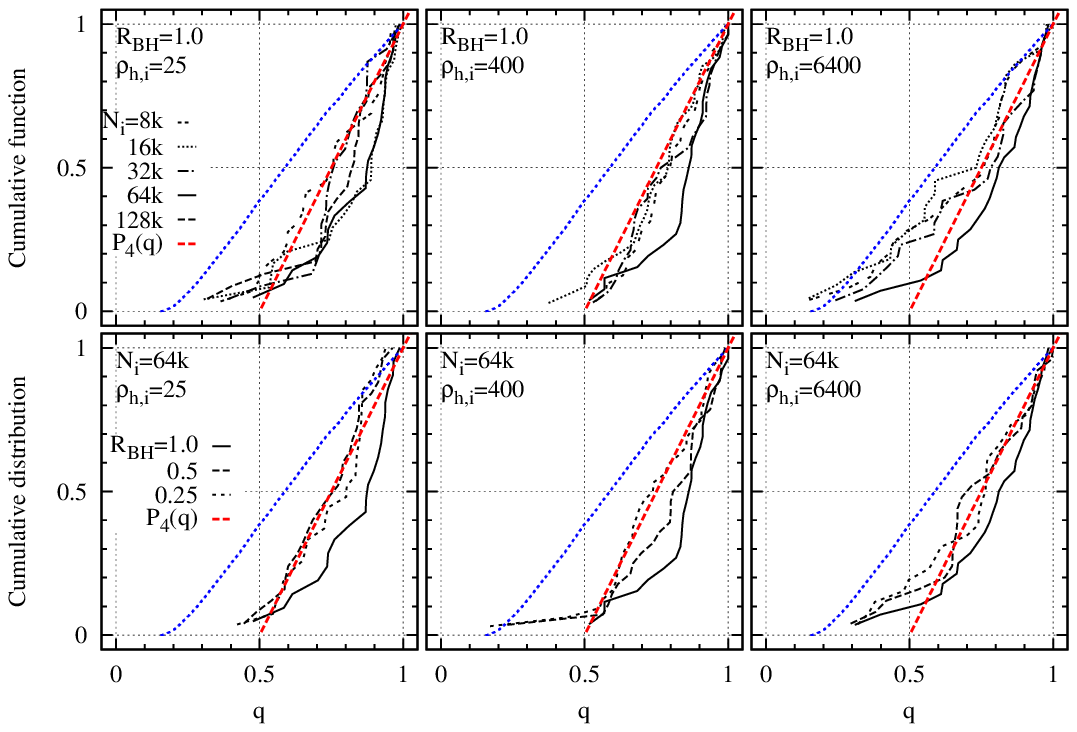}
 \end{center}
 \caption{Cumulative distribution of the mass ratios, $q=m_2/m_1$, of
   BBH escapers, where $m_2$ is the secondary masses of BBH
   escapers. Curves with the same line types indicate the same $N_{\rm
     i}$ among the top panels, and the same $R_{\rm BH}$ among the
   bottom panels. Red dashed and blue dotted curves are drawn as
   described in the main text.}
 \label{fig:cf_q}
\end{figure*}

We show cumulative distributions of mass ratios, $q=m_2/m_1$, of BBH
escapers in Fig.~\ref{fig:cf_q}, where $m_2$ is the secondary BH mass
of a BBH escaper. Blue dotted curves show the cumulative distribution
of the mass ratios of the artificially generated BBHs. Clearly, the
blue dotted curves are not at all in agreement with the cumulative
distribution in any simulation model. The distributions of the mass
ratios rather fit to the following function:
\begin{equation}
  P_4(q) = \int_0^q p_4(q') dq',
\end{equation}
where
\begin{equation}
  p_4(q) = 2 \;\; (0.5 < q < 1), \label{eq:fit_q}
\end{equation}
which are indicated by red dashed curves in each panel of
Fig.~\ref{fig:cf_q}. The mass ratio distributions of BBH escapers in
our simulations is larger than those of the artificially generated
BBHs. This is because more massive BHs are more likely to be at the
cluster centre, and to be retained in BBHs after binary-single and
binary-binary interactions. In fact, BBHs contain those with
$q<0.5$. However, such BBHs are at most $20$ per cent of all the BBHs
(simulation models with $\rho_{\rm h,i}=6400$). This fitting function
is sufficiently accurate for rough estimates of BBH mergers described
in section~\ref{sec:estimate}. We note that equation~(\ref{eq:fit_q})
is just a simple fitting. We do not claim that there is any physical
background for this probability distribution.

We mention some properties of NB-BBH escapers in simulation models
with $\rho_{\rm h,i}=6400$ and $R_{\rm
  BH}=1.0$. Fig.~\ref{fig:cf_cmev} summarises their properties. Their
cumulative distributions of eccentricities, primary masses, and mass
ratios fit to each fitting function (see the top right, the bottom
left, and the bottom middle panels, respectively). The time evolution
of the cumulative number of NB-BBH escapers is also in good agreement
with its fitting function for model with $N_{\rm i}=64k$ (see the
bottom right). This is natural, since the clusters of this model do
not include CE-BBH escapers.

\begin{figure*}
 \begin{center}
   \includegraphics[scale=1.4]{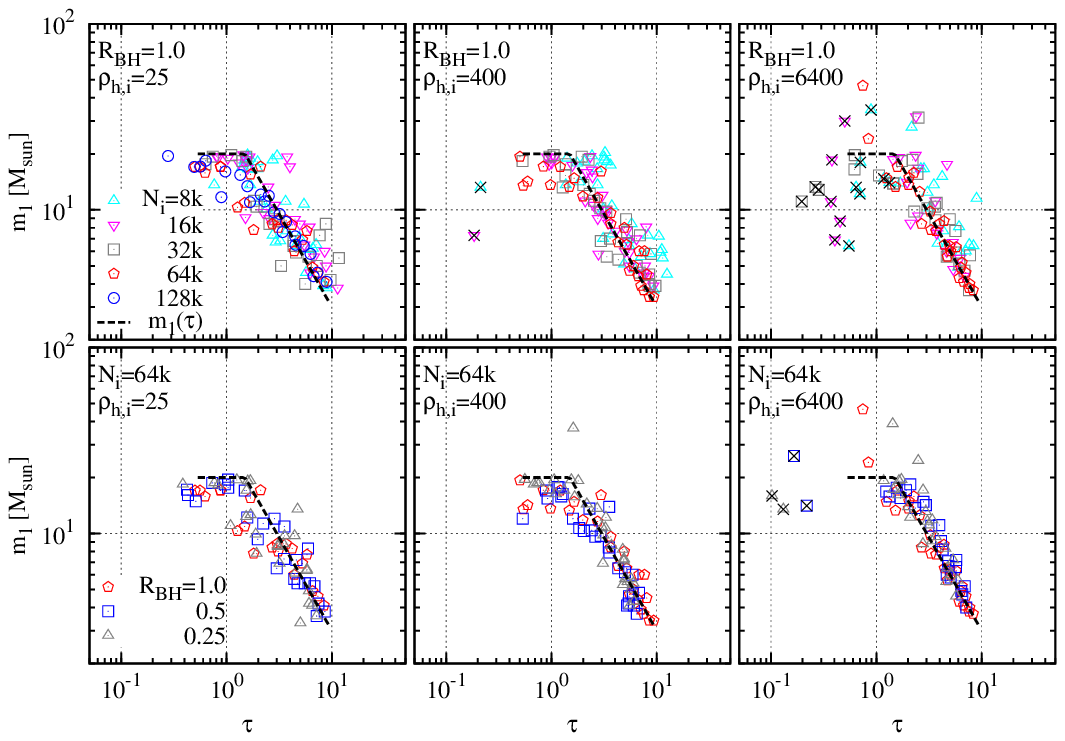}
 \end{center}
 \caption{Primary masses of BBHs which escape from a cluster at
   thermodynamical time, $\tau$. Points with the same point types
   indicate the same $N_{\rm i}$ among the top panels, and the same
   $R_{\rm BH}$ among the bottom panels. BBHs marked by black crosses
   are CE-BBHs.}
 \label{fig:tau_m1}
\end{figure*}

Finally, we search for correlations between BBH properties and BBH
escape time. Fig.~\ref{fig:tau_m1} shows primary masses of BBHs which
escape from clusters at $\tau$. Apart from CE-BBHs, the primary masses
become lighter with time in all the models in a similar way. Before
$\tau=1$, primary masses are about $20M_{\odot}$, which is maximum
mass of BHs without progenitor's mergers. The mass decreases
monotonically, and finally becomes about $3$ or $4M_{\odot}$ at
$\tau=10$. CE-BBHs escape at an early time, before $\tau=1$. They are
formed at an early time, and have large binding energies at
birth. Therefore, they are ejected from clusters once they interact
with other stars. Some of BHs have more than $20M_{\odot}$, and are
formed through mergers of their progenitors. No pair of BHs merges in
our simulations. As a simple fitting, we write the evolution of the
primary masses of BBH escapers as
\begin{align}
  m_1(\tau) = \left\{
  \begin{array}{ll}
    20 M_{\odot} & (0.5 < \tau < 1.5) \\ & \\ \displaystyle 20
    M_{\odot} \left( \frac{\tau}{1.5} \right)^{-1} & (\tau > 1.5)
  \end{array}
  \right., \label{eq:tau_m1}
\end{align}
which is indicated by black curves in Fig.~\ref{fig:tau_m1}.

Except the correlation between escape time and primary BH masses, we
have found no correlation between any pair of escape time, semi-major
axes, eccentricities, primary BH masses, and mass ratios.

\section{Estimate for BBH detection rate}
\label{sec:estimate}

In this section, we estimate a detection rate of GWs from mergers of
BBHs originating from GCs. In section~\ref{sec:formulation}, we give a
formula for a BBH detection rate, which is similar to those of O06 and
D10. Before we solve the formula, we need to adopt parameters related
to GCs, which is described in section~\ref{sec:GCmodel}. We describe
our solution method for the formula in
section~\ref{sec:solutionmethod}. Finally, we show the detection rate
and properties of BBH escapers in section~\ref{sec:properties}.

\subsection{Formula for BBH detection rate}
\label{sec:formulation}

In this section, we estimate a detection rate of BBHs originating from
GCs. We consider only NB-BBH escapers as GW sources. BBH escapers
merge more easily than BBHs inside GCs. Furthermore, NB-BBH escapers
dominate BBH escapers. This reason is described in
section~\ref{sec:GCmodel}.

We follow the approach of O06 and D10 for obtaining the detection
rate. The detection rate is calculated as
\begin{equation}
  \Gamma_{\rm det,tot} = f_{\rm det}^{-3} \int_{t_{\rm u}=0}^{t_{\rm
      u}=t_{\rm u,0}} \left[ n_{\rm gc} dV(t_{\rm u})
    \frac{\Gamma_{\rm det}(t_{\rm u})}{1+z(t_{\rm u})}
    \right], \label{eq:rdet,tot}
\end{equation}
which corresponds to equation~(13) of O06, and to equation~(11) of
D10, although their equations have been already discretised.  A
variable $t_{\rm u}$ is the universe age, and $t_{\rm u,0}$ is the
universe age at the present time. A variable $z(t_{\rm u})$ is a
redshift at a universe age $t_{\rm u}$. A variable $V(t_{\rm u})$ is
the volume of the universe at a universe age $t_{\rm u}$, which is
observed from us. A variable $\Gamma_{\rm det}(t_{\rm u})$ is a
detection rate of BBH escapers merging at the universe age $t_{\rm
  u}$, where the BBH escapers originate from {\it one} GC. A constant
$n_{\rm gc}$ is the number density of GCs in the universe. We adopt
$n_{\rm gc}=8.4h^3\mbox{Mpc}^{-3}$ \citep{PortegiesZwart00}, where $h$
is described later. A factor $f_{\rm det}=2.26$ considers the
non-uniform pattern of detector sensitivity and random sky orientation
of sources \citep{Finn93}. The factor $1/[1+z(t_{\rm u})]$ comes from
a cosmological time dilation of the detection rate. Here, we assume
that all GCs are identically formed at the same time, and that the
number density of GCs keeps constant at a given universe age $t_{\rm
  u}$.

We relate a universe age $t_{\rm u}$ to its redshift $z(t_{\rm u})$ as
\begin{align}
  t_{\rm u} &= H_0^{-1} \int_z^{\infty}
  \frac{dz'}{(1+z')\sqrt{\Omega_{\rm m}(1+z')^3+\Omega_{\Lambda}}}
  \\ &= \frac{1}{3H_0\sqrt{\Omega_{\Lambda}}} \log \left[
    \frac{\sqrt{\Omega_{\rm m}(1+z)^3 + \Omega_{\Lambda}} +
      \sqrt{\Omega_{\Lambda}}}{\sqrt{\Omega_{\rm m}(1+z)^3 +
        \Omega_{\Lambda}} - \sqrt{\Omega_{\Lambda}}} \right],
\end{align}
where $H_0=100h$~kms$^{-1}$Mpc$^{-1}$. We adopt $\Omega_{\rm m}=0.28$,
$\Omega_{\Lambda}=0.72$, and $h=0.73$, based on WMAP9
\citep{Hinshaw12}.

The volume $dV(t_{\rm u})$ is expressed as
\begin{equation}
  dV(t_{\rm u}) = 4 \pi D_{\rm p}^2 dD_{\rm p},
\end{equation}
where $D_{\rm p}$ is a proper distance. The proper distance is
expressed as
\begin{align}
  D_{\rm P} = \int_{t_{\rm u}}^{t_{\rm u,0}} \left[ 1 + z(t_{\rm u})
    \right] dt_{\rm u}.
\end{align}

We calculate $\Gamma_{\rm det}(t_{\rm u})$ as
\begin{equation}
  \Gamma_{\rm det}(t_{\rm u}) = \int_{C_{\rm det}>1} dm_1dq
  \frac{d\Gamma_{\rm mrg}(t_{\rm u})}{dm_1dq}. \label{eq:rdet}
\end{equation}
A variable $\Gamma_{\rm mrg}(t_{\rm u})$ is a merger rate of BBH
escapers at a universe age $t_{\rm u}$, where these BBH escapers
originate from one GC. A variable $C_{\rm det}$ indicates a
detectability of BBH escapers, and depends on $m_1$, $q$, $t_{\rm u}$
(i.e. a distance between an observer and a source), and a GW
observatory. Note that BBH escapers can be detected if $C_{\rm
  det}>1$. The merger rate $\Gamma_{\rm mrg}(t_{\rm u})$ is given by
\begin{equation}
  \Gamma_{\rm mrg}(t_{\rm u}) = \int_{t'_{\rm u}=t_{\rm u,0}-t_{\rm
      0}}^{t'_{\rm u}=t_{\rm u}} dN_{\rm BBH,esc}(t'_{\rm u}) D(t_{\rm
    u}-t'_{\rm u}), \label{eq:rmrg1}
\end{equation}
where $t_0$ is the GC age at the present time, $N_{\rm BBH,esc}(t_{\rm
  u})$ is the total number of BBH escapers from one GC at the universe
age $t_{\rm u}$, and $D(t_{\rm u}-t'_{\rm u})$ is the distribution of
time delay between the time when BBHs escape and the time when the
BBHs merge. Using equation~(\ref{eq:fit_bbhesc}), we can rewrite
equation~(\ref{eq:rmrg1}) as
\begin{align}
  \Gamma_{\rm mrg}(t_{\rm u}) = &0.020N_{\rm BH,i} \nonumber
  \\ &\times \int_{t_{\rm u,0}-t_0}^{t_{\rm u}} dt'_{\rm u} D(t_{\rm
    u}-t'_{\rm u}) \left. \frac{d\tau}{dt} \right|_{t=t'_{\rm
      u}-(t_{\rm u,0}-t_0)}, \label{eq:rmrg2}
\end{align}
where
\begin{align}
  N_{\rm BH,i} = 190 R_{\rm BH} \left( \frac{N_{\rm i}}{128k}
  \right). \label{eq:NBHi}
\end{align}
The derivative $d\tau/dt$ is obtained in
section~\ref{sec:solutionmethod}.

BBH escapers merge through GW radiation. Their merging timescales are
estimated as
\begin{align}
  t_{\rm GW} = \frac{5}{256} \frac{c^5}{G^3} \frac{a^4}{m_1^3 q(1+q)}
  g(e), \label{eq:tgw}
\end{align}
where
\begin{align}
  g(e) = \frac{(1-e^2)^{3.5}}{1+(73/24)e^2+(37/96)e^4}.
\end{align}
The constant $c$ is the light speed. Distribution function of BBH
escaper properties, $p(a,e,m_1,q)$, is approximated as
follows. Primary masses of BBH escapers, $m_1$, depends on a
thermodynamical time $\tau$ as equation~(\ref{eq:tau_m1}) (see also
Fig.~\ref{fig:tau_m1}), which can be transformed to a physical time
$t$. On the other hand, $a$, $e$, and $q$ are independent of
$t$. Therefore, we use a distribution function of BBH escaper
properties, $p(a,e,q)$ at a given universe time $t_{\rm
  u}$. Furthermore, $a$, $e$, and $q$ are uncorrelated with each
other. The distribution function $p(a,e,q)$ can be simplified as
\begin{align}
  p(a,e,q) = p_1(a) p_2(e) p_4(q). \label{eq:paeq}
\end{align}

The detectability of BBH escapers, $C_{\rm det}$, is given by
\begin{align}
  C_{\rm det} = \left( \frac{D_{\rm L}}{D_{\rm L,0}} \right)^{-1}
  \left( \frac{\mathcal{M}_{\rm ch}}{\mathcal{M}_{\rm ch,0}}
  \right)^{5/6} \sqrt{\frac{s(f_{\rm off})}{s(f_{\rm off,0})}},
\end{align}
where $D_{\rm L}$ $(=[1+z(t_{\rm u})]D_{\rm P})$ is a luminosity
distance, $\mathcal{M}_{\rm ch}$ is a redshifted chirp mass, and
$s(f_{\rm off})$ is a detector response function. The redshifted chirp
mass is expressed as
\begin{align}
  \mathcal{M}_{\rm ch} = \left[ 1 + z(t_{\rm u}) \right] m_{\rm ch},
\end{align}
where $m_{\rm ch}$ is a chirp mass, given by
\begin{align}
  m_{\rm ch} = \frac{q^{3/5}}{(1+q)^{1/5}} m_1.
\end{align}
The detector response function is approximated as
\begin{align}
  s(f_{\rm off}) = \int_0^{f_{\rm off}} \frac{(f')^{-7/3}}{S_{\rm
      N}(f')} df',
\end{align}
where
\begin{align}
  S_{\rm N}(f) \propto \left\{
  \begin{array}{ll}
    \infty & (f < 10 \mbox{Hz}) \\ \left( \frac{f}{f_0} \right)^{-4} +
    2 \left[ 1+ \left( \frac{f}{f_0} \right)^2 \right] & (f \ge 10
    \mbox{Hz})
  \end{array}
  \right.,
\end{align}
with $f_0=70$~Hz. The cut-off frequency, $f_{\rm off}$, is
approximated as
\begin{align}
  \left( \frac{f_{\rm off}}{\mbox{Hz}} \right) \sim 200 \left[
    \frac{m_1(1+q)}{20M_{\odot}} \right]^{-1} \left[
    \frac{1}{1+z(t_{\rm u})} \right].
\end{align}
In order to determine $D_{\rm L,0}$, $\mathcal{M}_{\rm ch,0}$, and
$s(f_{\rm off,0})$, we adopt that initial LIGO and next-generation GW
observatories can detect a merger of two NSs at distances of
$18.4$~Mpc and $300$~Mpc, respectively.

\subsection{Analysis models of GCs}
\label{sec:GCmodel}

In order to solve equation~(\ref{eq:rmrg2}), we need to set up
analysis models with various of $N_{\rm i}$, $\rho_{\rm h,i}$, and
$t_0$. A GC age $t_0$ is explicitly included in
equation~(\ref{eq:rmrg2}). The number of BHs at the initial time,
$N_{\rm BH,i}$, depends on $N_{\rm i}$, as seen in
equation~(\ref{eq:NBHi}). The time delay distribution $D(t_{\rm
  u}-t'_{\rm u})$ depends on semi-major axes of BBH escapers, $a$, via
merging timescales through GW radiation (see
equation~(\ref{eq:tgw})). As seen in section~\ref{sec:propBBH},
semi-major axis distribution of BBH escapers is independent of $N_{\rm
  i}$ and $\rho_{\rm h,i}$ in the unit of $a_{\rm 1kT_{\rm i}}$, which
depends on $N_{\rm i}$ and $\rho_{\rm h,i}$ (see
equation~(\ref{eq:abkt})). Therefore, $D(t_{\rm u}-t'_{\rm u})$ is
dependent on $N_{\rm i}$ and $\rho_{\rm h,i}$.

\begin{table}
  \caption{Analysis models for estimates of GW detection rates.}
  \label{tab:gcmodel}
  \begin{center}
    \begin{tabular}{cccc}
      \hline
      model & $N_{\rm i}$ & $\rho_{h,i}/M_{\odot}\mbox{pc}^{-3}$ & $t_0/\mbox{Gyr}$ \\
      \hline
      standard               & $1 \times 10^6$ & $1.0 \times 10^5$ & $10$ \\
                             & $1 \times 10^6$ & $1.0 \times 10^5$ & $12$ \\
      small-$N_{\rm i}$      & $5 \times 10^5$ & $1.0 \times 10^5$ & $10$ \\
                             & $5 \times 10^5$ & $1.0 \times 10^5$ & $12$ \\
      large-$N_{\rm i}$      & $2 \times 10^6$ & $1.0 \times 10^5$ & $10$ \\
                             & $2 \times 10^6$ & $1.0 \times 10^5$ & $12$ \\
      small-$\rho_{\rm h,i}$ & $1 \times 10^6$ & $6.4 \times 10^3$ & $10$ \\
                             & $1 \times 10^6$ & $6.4 \times 10^3$ & $12$ \\
      large-$\rho_{\rm h,i}$ & $1 \times 10^6$ & $1.0 \times 10^6$ & $10$ \\
                             & $1 \times 10^6$ & $1.0 \times 10^6$ & $12$ \\
      \hline
    \end{tabular}
  \end{center}
\end{table}

Our choice for $N_{\rm i}$, $\rho_{\rm h,i}$, and $t_0$ is summarised
in Table~\ref{tab:gcmodel}. These $t_0$ are in good agreement with
galactic GCs. These $N_{\rm i}$ and $\rho_{\rm h,i}$ are consistent
with initial conditions of M~4 \citep{Heggie08}, NGC~6397
\citep{Giersz09}, and 47~Tuc \citep{Giersz11}. Hereafter, we
collectively refer to these models as ``analysis models''.

In these analysis models, NB-BBH escapers will dominate BBH
escapers. The large-$N_{\rm i}$ model has the smallest half-mass
relaxation time among these models. The half-mass relaxation time of
the large-$N_{\rm i}$ model is almost the same as that of simulation
models with $N_{\rm i}=64k$ and $\rho_{\rm h,i}=6400$, in which the
fraction of CE-BBH escapers is $10$ per cent of the total BBH escapers
(see the tenth column of Table~\ref{tab:model}). Since the number of
CE-BBH escapers depends on half-mass relaxation time, the fraction of
CE-BBH escapers will be only $10$ per cent. CE-BBH escapers in the
other analysis models should be smaller than those in the
large-$\rho_{\rm h,i}$ model.

\subsection{Solution method}
\label{sec:solutionmethod}

\begin{figure*}
 \begin{center}
   \includegraphics[scale=1.0]{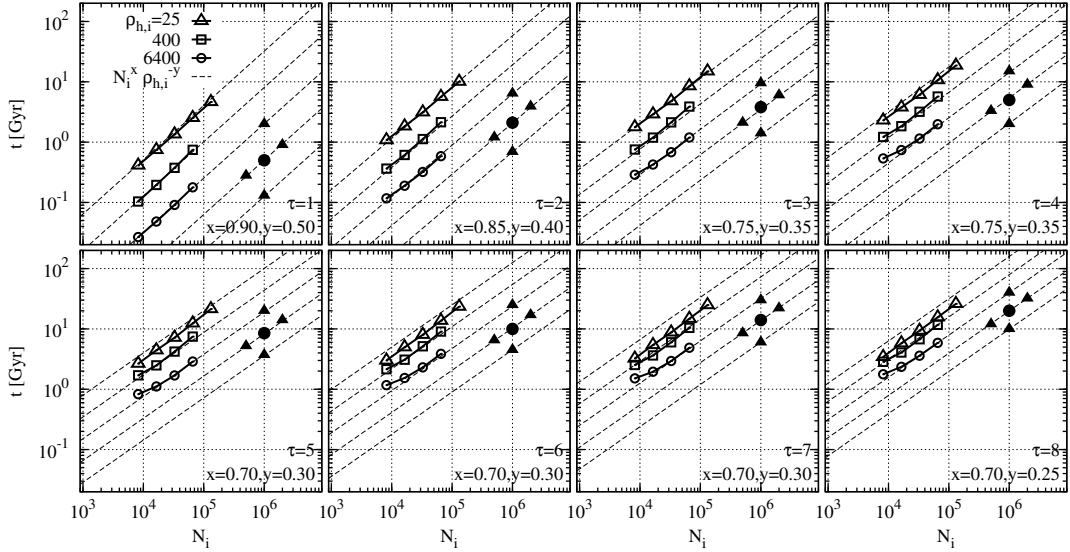}
 \end{center}
 \caption{GC ages at thermodynamical time $\tau=1$, $2$, $3$, $4$,
   $5$, $6$, $7$, and $8$ in our cluster models. Dashed lines fit to
   our simulation results, indicated by open points with solid
   curves. Fitting formula for these dashed lines are $N_{\rm
     i}^x\rho_{\rm h,i}^{-y}$, where the powers $x$ and $y$ is written in
   the panels. Filled points indicate physical time in the standard
   (circles) and the other models (triangles), which are obtained from
   the fitting formula.}
 \label{fig:tau_param}
\end{figure*}

We construct fitting formula for $\tau$ as a function of $t$ to obtain
$d\tau/dt$ in equation~(\ref{eq:rmrg2}) as
follows. Fig.~\ref{fig:tau_param} shows $t$ at $\tau$ in models with
various $N_{\rm i}$ and $\rho_{\rm h,i}$. We fit formula $N_{\rm i}^x
\rho_{\rm h,i}^{-y}$ for $t$ at each $\tau$, where the powers $x$ and
$y$ are indicated in Fig.~\ref{fig:tau_param}. For this fitting, we
ignore $\rho_{\rm h,i}=25$ models for $t$ in the cases of $\tau \ge
4$, which is justified as follows. In models with $\rho_{\rm h,i}=25$,
$\tau$ increases, compared with $t$. This is because the number of
stars, $N$, is decreased by evaporation due to strong tidal fields. On
the other hand, clusters in models with $\rho_{\rm h,i}=400$ and
$6400$ are not evaporated so much, which is the case for the analysis
models in Table~\ref{tab:gcmodel}. In order to obtain the relation
between $t$ and $\tau$, we can ignore the relation between $t$ and
$\tau$ in models with $\rho_{\rm h,i}=25$.

Note that we may overestimate $t$ at a given $\tau$ in the analysis
models, especially at large $\tau$. As indicated in
Fig.~\ref{fig:tau_param}, the power of $\rho_{\rm h,i}$, $y$,
decreases with $\tau$ increasing. This reflects the difference of
evaporation between models with $\rho_{\rm h,i}=400$ and
$6400$. Clusters in simulation models with $\rho_{\rm h,i}=6400$ model
are evaporated less than those with $\rho_{\rm h,i}=400$. The former
clusters keep their half-mass relaxation time larger, compared with
the latter clusters. If $\rho_{\rm h,i}$ is sufficiently large,
clusters are not evaporated regardless of their $\rho_{\rm h,i}$. The
power of $\rho_{\rm h,i}$, $y$, should be $0.5$, since half-mass
relaxation time is proportional to $\rho_{\rm h,i}^{-0.5}$ (see
equation~(\ref{eq:relaxationtime})). We may overestimate $t$ in
analysis models with $\rho_{\rm h,i}=10^5$ and $10^6$. We discuss this
in section~\ref{sec:uncertainty}.

\begin{figure}
 \begin{center}
   \includegraphics[scale=1.5]{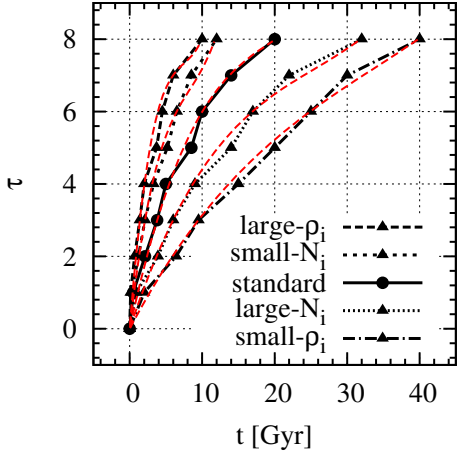}
 \end{center}
 \caption{Thermodynamical time $\tau$ at each physical time $t$ in the
   standard (filled circles) and the other models (filled
   triangles). Dashed red curves are obtained from third polynomial
   interpolation, and the fitting formulae is shown in
   Table~\ref{tab:tau_tphy}.}
 \label{fig:tau_tphy}
\end{figure}

From the formula $N_{\rm i}^x \rho_{\rm h,i}^{-y}$, we obtain $t$ of
the analysis models. Fig.~\ref{fig:tau_tphy} shows $\tau$ as a
function of $t$ in all the analysis models. For each model, we adopt
fitting formulae, using third polynomial interpolations. These fitting
formulae are summarised in Table~\ref{tab:tau_tphy}.

\begin{table}
  \caption{Fitting formula for $\tau=at^3+bt^2+ct$}
  \label{tab:tau_tphy}
  \begin{center}
    \begin{tabular}{cccc}
      \hline
      model & a & b & c \\
      \hline
      standard             & $6.52\times10^{-4}$ & $-3.96\times10^{-2}$ & $9.30\times10^{-1}$ \\
      small-$N_{\rm i}$    & $6.98\times10^{-3}$ & $-1.74\times10^{-1}$ & $1.76\times10^{+0}$ \\
      large-$N_{\rm i}$    & $2.50\times10^{-4}$ & $-1.91\times10^{-2}$ & $6.06\times10^{-1}$ \\
      small-$\rho_{\rm i}$ & $7.29\times10^{-5}$ & $-7.40\times10^{-3}$ & $3.80\times10^{-1}$ \\
      large-$\rho_{\rm i}$ & $1.40\times10^{-2}$ & $-3.16\times10^{-1}$ & $2.56\times10^{+0}$ \\
      \hline
    \end{tabular}
  \end{center}
\end{table}

In order to calculate equation~(\ref{eq:rdet,tot}) and
(\ref{eq:rmrg2}), we discretise the integrals in these equations. We
set time bins to $1$~Gyr. These time bins seem too large, however are
sufficiently accurate for this estimate. This is because we have
already extrapolated the number of BBH escapers, and used rough
fitting formulae for $a$, $e$, $m_1$, and $q$ of BBH escapers.

When we calculate the time delay distribution $D(t_{\rm u}-t'_{\rm
  u})$ at each time bin in equation~(\ref{eq:rmrg2}), we generate
$10000$ BBH escapers artificially as follows. We determine $a$, $e$,
and $q$ of a BBH escaper by means of Monte Carlo technique. The
probability distribution of $a$, $e$, and $q$ is obeyed by
equation~(\ref{eq:paeq}). Furthermore, $m_1$ is set by the universe
age $t_{\rm u}$. Here, the universe age $t_{\rm u}$ is determined as
follows. We can derive $\tau_{\rm min}$ and $\tau_{\rm max}$ from the
time bin. We obtain $\tau$ in a normal distribution between $\tau_{\rm
  min}$ and $\tau_{\rm max}$ by means of Monte Carlo
technique. Finally, we convert $\tau$ to $t_{\rm u}$, and obtain
$m_1$. The generated BBH escapers are reutilised when
equation~(\ref{eq:rdet}) is calculated.

\subsection{Detection Rate and properties of BBH escapers}
\label{sec:properties}

\begin{figure}
 \begin{center}
   \includegraphics[scale=1.2]{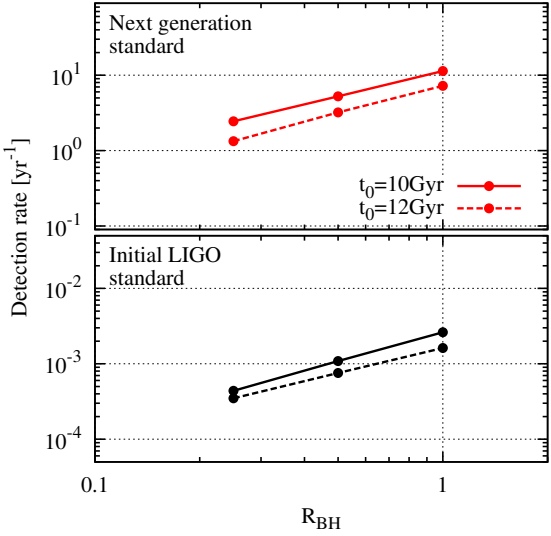}
 \end{center}
 \caption{Detection rates as a function of BH retention fraction
   $R_{\rm BH}$ in the standard model by means of next-generation GW
   observatories (top) and Initial LIGO (bottom). In each panel, the
   detection rates are indicated by solid and dashed curves, when the
   ages of GCs, $t_0$, are $10$~Gyr and $12$~Gyr, respectively.}
 \label{fig:rdet_std}
\end{figure}

Fig.~\ref{fig:rdet_std} shows detection rates by means of
next-generation GW observatories (top) and Initial LIGO (bottom) in
the standard model. The detection rate is much less than
$10^{-2}$~yr$^{-1}$ by means of Initial LIGO, while it is more than
$1$~yr$^{-1}$ by means of next-generation GW observatories. The
detection rates in younger analysis models ($t_0=10$~Gyr) are larger
than those in older analysis models ($t_0=12$~Gyr) by a factor of
about two (see the top panel of Fig.~\ref{fig:rdet_std}). More massive
BBHs escape, and merge at an earlier time, i.e. at a more distant
place. Therefore, the detection rate is larger in younger analysis
models.

\begin{figure*}
 \begin{center}
   \includegraphics[scale=1.0]{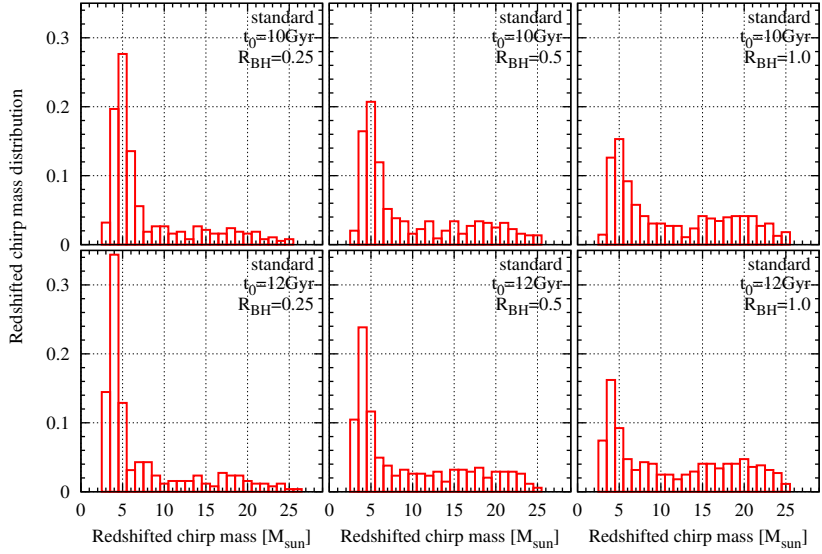}
 \end{center}
 \caption{Redshifted chirp mass distribution of detected BBHs in the
   standard model.}
 \label{fig:dmch_std}
\end{figure*}

Fig.~\ref{fig:dmch_std} shows redshifted chirp mass distributions in
the standard models. In each case of $t_0$ and $R_{\rm BH}$, its chirp
mass distribution has a sharp peak at about $5M_{\odot}$. This is
because our simulation models have a BH mass function with a peak
around at $5M_{\odot}$ (see Fig.~\ref{fig:bhmassfunc}), and because
less massive BBHs merge at a later time, i.e. at a closer
place. Comparing the chirp mass distributions between younger and
older analysis models, we can see that a chirp mass distribution has a
peak at more massive chirp mass in younger analysis models in each
$R_{\rm BH}$ model; younger and older GCs have a peak at $5M_{\odot}$
and $4M_{\odot}$, respectively. This is consistent with the above
argument that more massive BBHs merge in younger analysis models at a
given universe time (or at a given place).  The peaks in younger and
older analysis models becomes sharper as $R_{\rm BH}$ decreases. More
massive BBHs merge at an earlier time in smaller $R_{\rm BH}$ models,
and are harder to be detected. This is because BBH escapers have
smaller semi-major axes in smaller $R_{\rm BH}$ models (see
Fig.~\ref{fig:cf_axis}), i.e. smaller merging timescales.

\begin{figure}
 \begin{center}
   \includegraphics[scale=1.0]{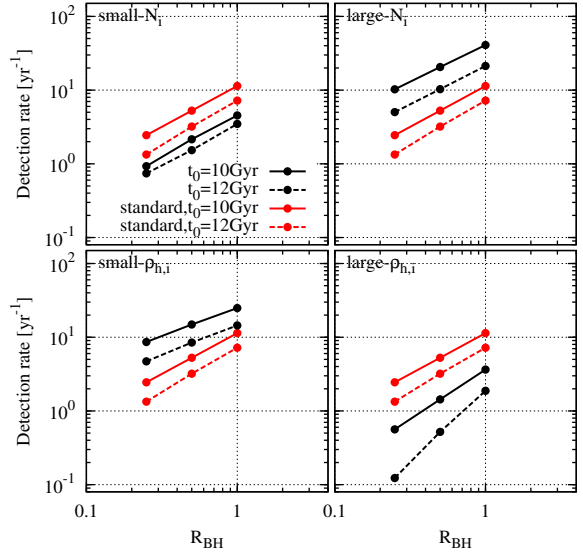}
 \end{center}
 \caption{Detection rates (black points with curves) as a function of
   BH retention fraction $R_{\rm BH}$ in models other than the
   standard model by means of next-generation GW observatories. For
   comparison, the detection rates in the standard model are indicated
   by red points with curves. In each panel, the detection rates are
   indicated by solid and dashed curves, when the ages of GCs, $t_0$,
   are $10$~Gyr and $12$~Gyr, respectively.}
 \label{fig:rdet_sub}
\end{figure}

Fig.~\ref{fig:rdet_sub} shows detection rates in models other than the
standard models (black points with curves). The detection rates in the
small-$\rho_{\rm h,i}$ and large-$N_{\rm i}$ models are larger than
those in the standard models, while those in the large-$\rho_{\rm
  h,i}$ and small-$N_{\rm i}$ models are smaller. Half-mass relaxation
time in the former models is larger than that in the standard
models. Therefore, the clusters evolve more slowly, and more massive
BBHs escape, and merge at a later time (or at a closer
place). Eventually, the detection rates become larger than the
standard models. In the latter models, the opposite happens.

\begin{figure*}
 \begin{center}
   \includegraphics[scale=1.0]{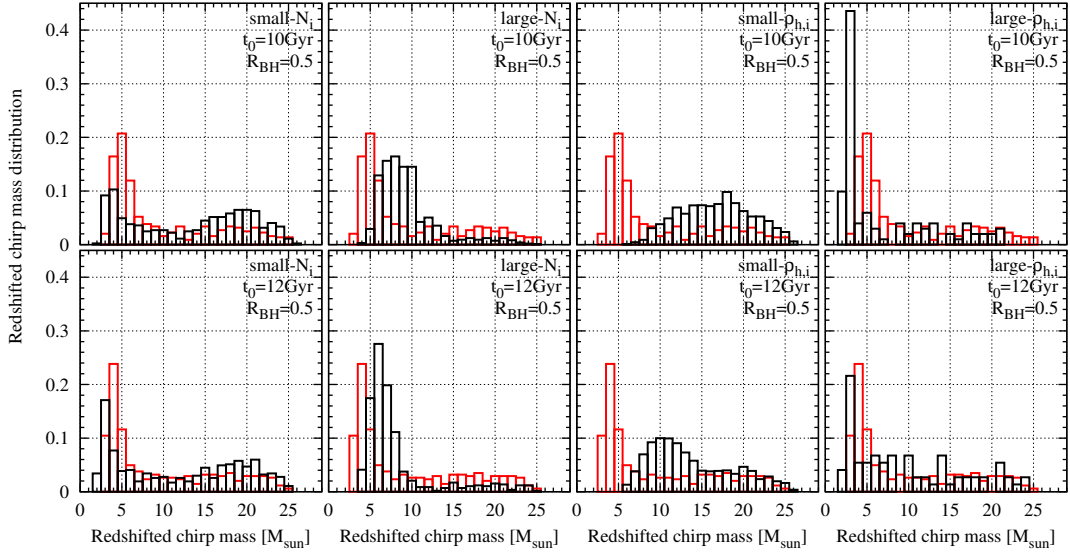}
 \end{center}
 \caption{Redshifted chirp mass distribution of detected BBHs in the
   standard (red) and the other models (black), where BH retention
   fraction $R_{\rm BH}=0.5$, and the age of GCs $t_0=10$~Gyr (top
   panels) and $t_0=12$~Gyr (bottom panels).}
 \label{fig:dmch_sub}
\end{figure*}

Fig.~\ref{fig:dmch_sub} shows redshifted chirp mass distribution of
merged BBH escapers in models other than the standard model. Chirp
mass distributions in the large-$N_{\rm i}$ and small-$\rho_{\rm h,i}$
models have peaks at larger masses than those in the standard
models. In these models, clusters evolve more slowly than in the
standard models. More massive BBHs escape, and merge at a later
time. Conversely, in the large-$\rho_{\rm h,i}$ models, the peak has
$3M_{\odot}$, which is smaller than in the standard models. In this
model, clusters evolve more rapidly, and more massive BBHs escape and
merge at an earlier time. In the small-$N_{\rm i}$ model, we can see
two peaks at $4M_{\odot}$ and $20M_{\odot}$ in $t_0=10$~Gyr case. The
smaller peak, which is smaller than that at the peak in the standard
model, can be explained by the reason similar to the above. In the
small-$N_{\rm i}$ model, clusters evolve more rapidly than in the
standard model, and more massive BBHs escape, and merge at an earlier
time. The larger peak comes from larger semi-major axes of BBH
escapers, i.e. their larger merging timescales, due to small $N_{\rm
  i}$. Since massive BBHs which escape at an early time have larger
merging timescales, the larger number of massive BBHs before merge is
left in this model than in the standard model.

\begin{figure}
 \begin{center}
   \includegraphics[scale=1.0]{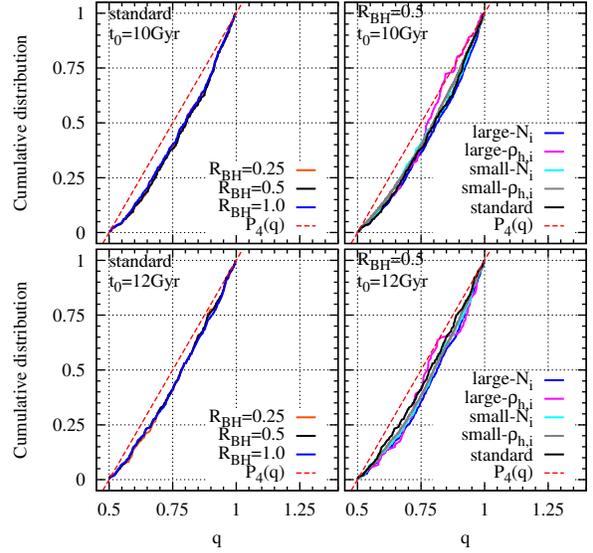}
 \end{center}
 \caption{Cumulative distribution of mass ratios of detected
   BBHs. Analysis models, GC ages, and BH retention fractions are
   indicated in each panel. Red dashed lines in all the panels are
   fitting formula for cumulative distribution of the mass ratios,
   which are the same as red dashed lines in Fig.~\ref{fig:cf_q}.}
 \label{fig:dmsr}
\end{figure}

We show cumulative distributions of mass ratios of detected BBHs in
Fig.~\ref{fig:dmsr}. The cumulative distributions are almost the same
regardless of analysis models. The mass ratios of detected BBHs shift
to unity, compared with those of the generated BBHs. This is because
BBHs with larger mass ratios are more likely to be
detected. Conservatively speaking, it may come from an artificial
effect that the mass ratios are almost the same. When we generate BBH
populations, we adopt the same probability distribution of the mass
ratios, whenever BBHs escape. This is because we do not find any
correlation between the mass ratios and physical time (or
thermodynamical time) in our simulations. However, the number of BBH
escapers in our simulations may be statistically poor in order to
reveal a correlation between the mass ratios and the time when BBHs
escape. Nevertheless, we emphasise the importance that the mass ratio
distribution is independent of analysis models, since this can be a
clue to make clear a dominant BBH formation process.

\section{Discussion}
\label{sec:discussion}

\subsection{Uncertainty in our estimate}
\label{sec:uncertainty}

In previous sections, we obtain a detection rate of BBH mergers by GW
observatories by simplification in many points. In this section, we
discuss effects of these simplifications on our estimate.

First, our simulations only span one order of magnitude in $N_{\rm
  i}$, $10^4$ -- $10^5$. It is uncertain that the $N_{\rm i}$
dependence observed in our simulations continues to $N_{\rm i} \sim
10^6$. This uncertainty affects many quantities, such as the number of
BBH escapers, semi-major axes of BBH escapers, and the relation
between $t$ and $\tau$. We note that the uncertainty of $N$-dependence
is inherent in our estimate of BBH detection rates.

We also extrapolate $\rho_{\rm h,i}$ by using $\rho_{\rm h,i}$
dependence, and often ignore results in simulation models with
$\rho_{\rm h,i}=25$. It is not trivial to apply results in simulation
models with $\rho_{\rm h,i}=400$ and $6400$ to estimate of BBH
detection rate in analysis models with $\rho_{\rm h,i}>10^5$. Clusters
with $\rho_{\rm h,i}>10^5$ are less evaporated than those with
$\rho_{\rm h,i}=400$ and $6400$. This difference can affect $\rho_{\rm
  h,i}$ dependence of the number and semi-major axes of BBH
escapers. Considering the difference of the number and semi-major axes
between models with $\rho_{\rm h,i}=25$ and models with $\rho_{\rm
  h,i}=400$ and $6400$, we may underestimate the number of BBH
escapers, and overestimate the semi-major axes of BBH escapers. It is
unclear whether we overestimate or underestimate BBH detection
rate. BBH detection rate increases with the number of BBH escapers
increasing, however decreases with the semi-major axes of BBH escapers
decreasing, since BBH escapers merge at an earlier time, i.e. at a
more distant place.

As described in section~\ref{sec:solutionmethod}, we may overestimate
a physical time of an analysis model at a given thermodynamical time,
and underestimate a speed of its dynamical evolution. If this is
true, BBHs are formed, and escape from clusters at an earlier time
than we estimate in section~\ref{sec:properties}. Then, BBH escapers
merge at a more distant place, and are harder to be
detected. Eventually, the BBH detection rates are decreased, and the
peaks of chirp mass distribution shift smaller masses in all the
analysis models.

We model BH natal kick, such that BHs are ejected from a cluster with
a fixed probability, independently of BH masses. Since our initial
stellar mass function is top-light (see equation (\ref{eq:kroupa01})),
BH populations in our clusters have a top-light mass function, as seen
in Fig.~\ref{fig:bhmassfunc}. However, a BH mass function in reality
may not be top-light. \cite{Belczynski06} have published BH mass
functions in young stellar clusters. Their fig.~6 shows that BH mass
function has two peaks at $10$ -- $16M_{\odot}$, and at $22$ --
$26M_{\odot}$. There are few BHs with less than $10M_{\odot}$ in
contrast with our BH mass function. The difference between our and
their BH mass functions comes from models of BH natal kick. They have
adopted a model of BH natal kick velocity inversely proportional to BH
masses. Our BH mass function may contain more BHs with $<10M_{\odot}$
than in reality.

In our simulation models, we adopt metallicity $Z=0.001$, which is one
of two peaks in a metallicity distribution of GCs. The other peak is
at $Z=0.02$. As metallicity becomes larger, stellar wind becomes
strong, and BH mass becomes smaller. Nevertheless, our BH masses are
smaller than BH masses at $Z=0.02$ \citep{Belczynski06}.

We do not consider primordial binaries. The primordial binaries should
contain binaries consisting of two BH progenitors. Some of these
binaries should be compact enough to survive against external
perturbations from other cluster stars. They can experience common
envelop evolution, and become CE-BBHs, or single BHs because of BH
progenitor mergers. Eventually, our models may underestimate the
relative importance of CE-BBHs and single BHs whose progenitors
experience mergers. However, we mention that CE-BBHs are formed in our
simulations more easily than in reality. We include no natal kick in a
part of BHs. If these BHs form CE-BBHs, the CE-BBHs are not
disrupted. In reality, BHs should receive natal kick velocities less
than escape velocities of clusters. This natal kick may disrupt
CE-BBHs.

We also ignore primordial mass segregation. If we consider primordial
mass segregation, massive stars are initially concentrated at the
cluster centre. This effect should increase the number of massive
stars which experience common envelop evolution, and which merge with
another massive star.

BBH escapers in our simulation results slightly contain BBHs in
hierarchical triple systems. We have found such BBHs in a part of
cluster models; for example, one BBH of $28$ BBHs is included in
hierarchical triple systems in a cluster model with $N_{\rm i}=64k$,
$\rho_{\rm h,i}=6400$, and $R_{\rm BH}=1.0$. However, such BBHs can be
increased if primordial binaries are considered. This is because
binary-binary interactions leave hierarchical triple systems at a
significant rate \citep{Mikkola83,Mikkola84}. BBHs in hierarchical
triple systems are secularly perturbed by the third stars, and change
their eccentricities periodically \citep{Kozai62}. BBHs in
hierarchical triple systems have smaller merging timescales than those
in isolation
\citep{Miller02,Blaes02,Thompson11,Pejcha13,Seto13}. Therefore, we may
overestimate the merging timescales of BBH escapers.

In summary, our BH mass function may have too top-light shape as
compared to a BH mass function in reality. It is unclear that these
increase or decrease the BBH detection rate. More massive BBHs escape,
and merge at an earlier time. However, they are detected more easily
owing to larger chirp masses. BBHs in hierarchical triple systems also
merge at an earlier time, which may decrease our BBH detection rate.

So far, we ignore the detection of BBH mergers inside GCs. Our
simulations results show that some of BHs remain in GCs with the
standard, large-$N_{\rm i}$, and small-$\rho_{\rm h,ii}$ models. This
is consistent with a recent observation of \cite{Strader12}, who have
found in M22 two radio sources which seem BHs. These BHs should form
BBHs. Although BBHs in GCs have larger semi-major axes than BBH
escapers, they should be frequently perturbed by other stars, have
very high eccentricities by chance, and merge. In this sense, we may
underestimate the detection rate of BBH mergers.

\subsection{Comparison with previous studies}
\label{sec:comparison}

In this section, we compare our estimate with previous ones. It is
difficult to compare these estimates, since both of numerical methods
and simulation models, such as BH mass functions, are different. If we
ignore the difference of numerical methods, our estimate ($0.3$ --
$10$~yr$^{-1}$) is consistent with the estimates of O06 ($1$ --
$10$~yr$^{-1}$), and D10 ($1$ -- $100$~yr$^{-1}$), except that of S08
($25$ -- $3000$~yr$^{-1}$), although BH mass functions are completely
different. BBH detection rates may be weakly dependent on BH mass
functions. However, we need to confirm whether we get similar
estimates to those of O06 and D10 if we adopt the same BH mass
functions as theirs. This will be our future work.

\subsection{Distinction of BBH formation process}
\label{sec:distinction}

In this section, we discuss the possibility of distinction of a
dominant BBH formation process, such as formation from a primordial
binary on galactic fields, and dynamical formation in dense stellar
clusters. From the above, a detection rate of BBHs formed in GCs
should be at most $100$~yr$^{-1}$. This is similar to a detection rate
of BBHs originating from galactic fields \citep[e.g.]{Abadie10}. Note
that \cite{Belczynski12} have reported $10^3$ -- $10^4$~yr$^{-1}$. It
may be difficult for detection rates to make clear a dominant BBH
formation process.

A mass ratio distribution can be promising for its distinction. As
described in section~\ref{sec:properties}, the mass ratio distribution
is independent of analysis models.  However, we postpone trying to use
mass ratio distribution. Mass ratio distribution will strongly depend
on BH mass functions. In our future work, we will investigate
dependence of mass ratio distributions on BH mass functions, and
search for a method of the distinction of a dominant BBH formation
processes.

As a chirp mass distribution is different among analysis models, the
distribution may be useful to know GC initial conditions. Chirp mass
distribution will also depend on BH mass functions. In future work, we
will also seek its dependence.

\section{Summary}
\label{sec:summary}

We have performed $N$-body simulations in order to estimate the
detection rate of mergers of BBHs originating from GCs by means of GW
observatories. We have also obtained their distributions of chirp
masses and mass ratios. $N$-body simulations can not deal with a GC
with $N \sim 10^6$ due to the lack of computing power. Instead, we
have performed $N$-body simulations of small-$N$ clusters ($N_{\rm
  i}=8k$ -- $128k$), and have extrapolated their results to large-$N$
clusters.

This extrapolation can be done, using the following properties of BBH
escapers. BBHs are ejected from the clusters at a rate proportional to
$N_{\rm i}$ in the unit of $\tau$. The semi-major axis distribution of
BBH escapers are independent of $N_{\rm i}$ in the unit of $a_{\rm
  1kT_{\rm i}}$, i.e. inversely proportional to $N_{\rm i}$ in
physical units. The distributions of eccentricities, primary masses,
and mass ratios of BBH escapers are independent of $N_{\rm i}$. We
have estimated the detection rate of BBH mergers by the
next-generation GW observatories. The detection rate is $0.1$ --
$10$~yr$^{-1}$ for $R_{\rm BH}=0.25$, $0.5$ -- $20$~yr$^{-1}$ for
$R_{\rm BH}=0.5$, and $2$ -- $40$~yr$^{-1}$ for $R_{\rm BH}=1.0$. The
difference of the detection rates in each $R_{\rm BH}$ comes from
initial conditions of GCs, such as $N_{\rm i}$, $\rho_{\rm h,i}$, and
$t_0$.

Our estimate of a BBH detection rate is almost the same as previous
studies of BBHs in GCs (O06; D10; D11). Furthermore, it is similar
both to those of BBHs formed in galactic centres \citep{OLeary09}, and
to those of BBHs formed on galactic fields \citep{Abadie10}. These BBH
detection rates can not distinguish a dominant BBH formation process.

A mass ratio distribution of BBHs is independent of GC initial
conditions. It may be a clue to constrain a BBH formation process. In
future work, we will investigate the mass ratio distributions in
various type of BBH mass functions, and compare them with those on
galactic fields. Since a chirp mass distribution of BBHs is sensitive
to GC initial conditions, this distribution may make clear GC initial
conditions, if detected BBHs are dominantly formed in GCs.

\section*{Acknowledgements}

The author thanks Kohji Yoshikawa for discussion on GPU code, and
Keigo Nitadori for providing me with {\tt Yebisu} code. Numerical
simulations have been performed with HA-PACS at the Center for
Computational Sciences in University of Tsukuba. This work was in part
supported by Grant-in-Aid for Scientific Research (S) by JSPS
(20224002).

\bsp

\label{lastpage}

\end{document}